\documentclass[12pt,
reprint,
amsmath,
amssymb,
longbibliography,
aps,
prstab,
]{revtex4-1}
\usepackage[]{units}
\usepackage{graphicx}
\usepackage{dcolumn}
\usepackage{bm}
\usepackage{color}
\usepackage{soul} 

\begin{document}

\title{Collimation of partially stripped ions in the CERN Large Hadron Collider}

\newcommand{\lead}{$^{208}$Pb$^{82+}$~}
\newcommand{\leadpsi}{$^{208}$Pb$^{81+}$~}

\newcommand*{\affmark}[1][*]{\textsuperscript{#1}}
\author{A. Gorzawski\affmark[1,2]}
\thanks{arek.gorzawski@cern.ch}%
\author{A.~Abramov\affmark[1,3]}
\thanks{andrey.abramov@cern.ch}
\author{R.~Bruce\affmark[1]}
\author{N.~Fuster-Martínez\affmark[1]}
\author{M.~Krasny\affmark[1,4]}
\author{J.~Molson\affmark[1]}
\author{S.~Redaelli\affmark[1]}
\author{M.~Schaumann\affmark[1]}
\affiliation{\affmark[1]CERN European Organization for Nuclear Research,  Esplanade des Particules 1, 1211 Geneva, Switzerland, \\ 
			 \affmark[2]University of Malta, Msida, MSD 2080 Malta\\
			 \affmark[3]JAI, Egham, Surrey, United Kingdom\\
			 \affmark[4]LPNHE, Sorbonne University, CNRS/INP2P3, Tour 33, RdC, 4, pl. Jussieu, 75005 Paris, France\\}

\date{\today}

\begin{abstract}
In the scope of the Physics Beyond Colliders studies, the Gamma-Factory initiative proposes the use of partially stripped ions as a driver of a new type of high-intensity photon source in CERN Large Hadron Collider (LHC). In 2018, the LHC accelerated and stored partially stripped $^{208}\text{Pb}^{81+}$ ions for the first time. The collimation system efficiency recorded during this test was found to be prohibitively low, so that only a very low-intensity beam could be stored without the risk of triggering a beam dump when regular, minor beam losses occur. The worst losses were localised in the dispersion suppressor of the betatron-cleaning insertion. This article presents an analysis to understand in detail the source of these losses. Based on this understanding, possible mitigation measures that could significantly improve the cleaning efficiency and enable regular operation with partially-stripped ions in the future are developed. 

\end{abstract}
\keywords{partially stripped ions, LHC, high-intensity photon source}
\maketitle

\section{Introduction}
\label{sec:intro}

The CERN Large Hadron Collider (LHC) is designed to provide proton collisions at an energy of 7 TeV per beam and heavy-ion collisions at the equivalent magnetic rigidity~\cite{lhcdesignV1}. It consists of eight arcs and eight straight Insertion Regions (IRs). The accelerated beams collide in four out of the eight IRs, where the particle-physics experiments are located. The baseline design of the LHC includes operation with protons and lead ion beams (\lead), but other beam particle species have also been considered  \cite{YR_WG5_2018, Schaumann:IPAC2018-MOPMF039}.

The Physics Beyond Colliders initiative \cite{alemany2019summary} is a dedicated research campaign steered by CERN, focused on exploring alternative options to colliders for future particle-physics experiments. One of the projects under consideration 
is the Gamma Factory (GF), a design study for a novel type of light source \cite{krasny2015gamma}. The goal of the study group is to explore the possibilities to use the LHC beams for creating high-intensity, high-energy photon beams that can serve many applications, spanning from atomic to high-energy physics. 

The concept relies on using partially stripped ions (PSI) beams in the LHC as a driver. PSI retain one or more bound electrons. To produce photons, the remaining electrons in PSI are exited using a laser. The energy of the photons emitted during the spontaneous de-excitation of the excited atomic states is proportional to the square of the Lorentz factor of the ion beam, which allows photon energies of up to \unit[400]{MeV} in the LHC. While the proof-of-principle experiment for the GF  is proposed to operate at the Super Proton Synchrotron (SPS) at a beam energy of \unit[450]{Z GeV}~\cite{Krasny:2690736}, the ultimate  implementation is intended to operate at the LHC's top energy of \unit[7]{Z TeV}. 

The GF depends on the acceleration and storage of PSI beams in CERN's accelerator complex. PSI at varying states of ionisation are routinely used at CERN during the different stages of acceleration of the typical lead or argon beams in the injectors \cite{lhcdesignV3,kuchler14,Arduini:308372}. However, the LHC was never used to accelerate PSI beams.

In 2018, the first operational tests with PSI beams in the LHC were performed with the goal of studying the beam lifetime and characterising the beam losses for such a beams~\cite{Schaumann_2019_ipac_part_strip_ions,Schaumann:2670544}. During one dedicated machine development (MD) session, PSI beams were successfully injected, accelerated, and stored. At the same time, it was found  that the current LHC configuration poses a critical limitation on PSI operation. The collimation-system efficiency recorded during this test was found to be orders of magnitude worse than in standard operation and hence prohibitively low, effectively imposing an intensity limit. The worst losses were localised in the dispersion suppressor (DS) of the betatron-cleaning insertion. 
These findings 
clearly put in question the overall feasibility to operate the LHC with PSI beams of sufficient intensities for a future GF facility.

In this article we study the underlying physical processes responsible for the worsening in collimation efficiency with PSI beams and study possible mitigation measures.  In Section~\ref{sec:lhc-recap} we provide a brief recap of the LHC operation, collimation, and beam loss limitations. Section~\ref{sec:experiments} describes the experiments performed in the LHC using PSI beams, as well as the encountered limitations.  Section~\ref{sec:simulations} describes the interpretation of the observed losses with PSI beams and then describes the available simulation tools for the case of partially stripped ions. In the same section, the measured loss maps are compared against results from simulations. Finally, in Section \ref{sec:mitigations}, different mitigation strategies for the found limitations are outlined and investigated, including a new DS collimator, crystal collimation, or an orbit bump.


\section{LHC collimation and beam losses}
\label{sec:lhc-recap}

\subsection{LHC collimation system}

In the LHC, the proton-beam stored energy in Run~2 (2015-2018) at \unit[6.5]{TeV} exceeded \unit[300]{MJ}, approaching the design value of \unit[362]{MJ} planned for the operation at \unit[7]{TeV}. Uncontrolled beam losses of even a tiny fraction of the full beam could cause a superconducting magnet to quench or even cause material damage of exposed accelerator components. Therefore, a multi-stage collimation system is installed. It consists of more than 100 movable devices, installed to provide beam cleaning and passive protection against beam losses during regular operation and accidents~\cite{lhcdesignV1,assmann05chamonix,assmann06,bruce14_PRSTAB_sixtr,bruce15_PRSTAB_betaStar,valentino17_PRSTAB}. The betatron-halo cleaning is done by a three-stage collimator hierarchy in IR7 while off-momentum cleaning is done in IR3 by a similar system. Collimators for local triplet-magnet protection are located in all experimental IRs (IR1, IR2, IR5 and IR8) and, in addition, collimators for the physics-debris cleaning are installed in the high-luminosity experiments in IR1 and IR5.

\begin{figure}[t]
\centering
  \includegraphics[width=\columnwidth]{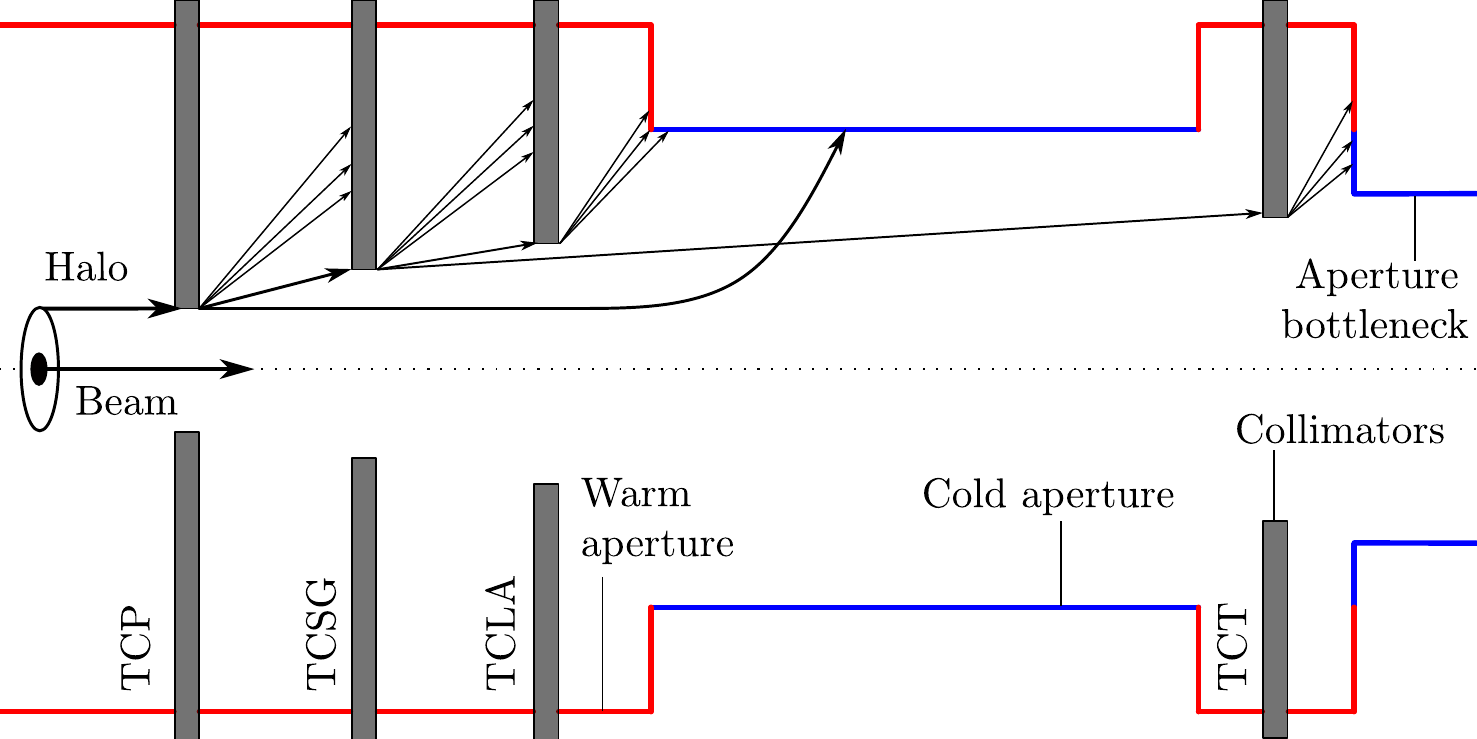}
\caption{Conceptual sketch of the layout of LHC collimators around the ring. Collimators are located mainly in IR3 and IR7, but also protect the experiments, the beam dump, and the injection regions.}
\label{fig:collimators-around-LHC}
\end{figure}

In the three-stage collimation system of the LHC, the primary collimators (TCP) are the devices closest to the beam of the whole machine and their purpose is to intercept any halo particles drifting out to large amplitudes~\cite{redaelli_coll}. Particles that are not absorbed by the TCPs, but scattered to larger amplitudes should be caught by the secondary collimators (TCS), which are designed to intercept the secondary beam halo. As a third stage that uses shower absorber collimators (TCLAs) is in place to intercept the tertiary halo and the products of hadronic showers, leaking from the TCSs. 
Additional tertiary collimators (TCT) are placed around the experimental insertions. Figure~\ref{fig:collimators-around-LHC} illustrates the collimator locations around the LHC. It is important to note that beam-halo particles interacting with the TCPs are not always deflected onto the TCSs; in some cases they can escape the collimation insertion and complete further revolutions around the machine before being disposed of at collimators. Beam particles with momentum offsets escaping the collimation section can be lost on the cold aperture in the dispersion suppressor immediately downstream, where the rising dispersion affects their trajectories~\cite{bruce14ipac_DS_coll}. The DS in IR7 is thus the main bottleneck for beam-halo losses in the LHC and the amount of local losses in the DS may impose limitations on the total achievable intensity, in particular for heavy-ion operation~\cite{epac2004,hermes16_ion_quench_test,hermes16_nim}

\subsection{Beam loss measurements and loss maps}

In order to prevent quenches and damage of sensitive equipment, beam loss monitors (BLMs) are installed around the LHC ring~\cite{blmSystem,holzer05,holzer08a}. They trigger a beam dump if the local losses exceed a pre-defined threshold. 
The LHC BLM system uses about 4000 ionization chambers, placed outside the cryostat on cold magnets and on other key equipment such as collimators. The BLM system provides a measurement of electromagnetic and hadronic showers resulting from nearby impacts of beam losses. They provide data at a time resolution down to \unit[40]{$\mu$s}, i.e. around half the beam revolution time.

\begin{figure*}[t]
\centering
\includegraphics[width=\textwidth]{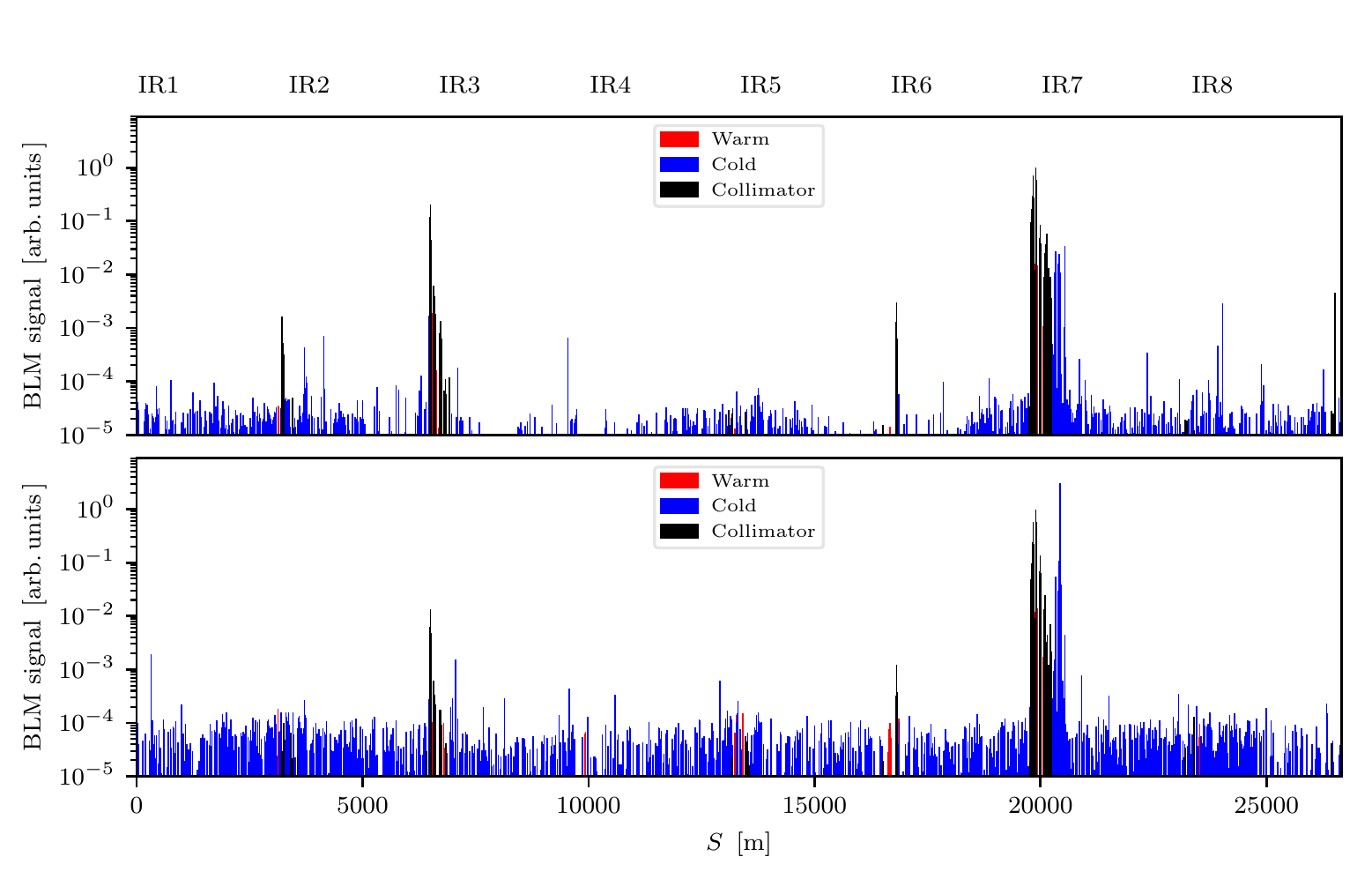}
\caption{LHC loss maps, indicating losses along the ring circumference. Black bars represent losses intercepted at the collimators. Blue regions represent the parts where superconducting magnets are located, and the red regions are related to the warm part of the ring. The top plot shows a loss map using a  regular  \lead ion beam  at \unit[6.37] {Z TeV} and the bottom plot shows a loss map with PSI (\leadpsi) taken at \unit[6.5] {Q TeV}, where $Z$ is an ion atomic number, $Q$ is the ion charge defined as $Q=Z-1$. }
\label{fig:example-loss-map}
\end{figure*}

In order to validate any operational configuration in the LHC, including the  optical configuration and the collimator positions, validation tests called loss maps (LMs) are performed. During these tests a safe, low-intensity beam is artificially excited to  create losses while the BLMs provide a continuous measurement of the loss distribution around the ring. An example of a loss map from standard operation with \lead ions can be seen in the top graph of Fig.\ref{fig:example-loss-map}.

Loss maps are carried out at all stages of the LHC operational cycle. Firstly, the beams are injected in the machine at \unit[450]{GeV}, at the so-called injection plateau. After the injection of bunches is finished, the energy of the two beams is increased for about 20 minutes during the energy ramp during which the first optics change occur. Once at top energy,  the optics is adjusted in order to achieve a lower $\beta$-function at the collision points, which is called the squeeze. Then, the beams are put in collision, and this part of operation is called physics.

\section{PSI beam tests in the LHC}
\label{sec:experiments}
\subsection{Experimental setup}

The collimation system in the LHC has been designed and optimised for proton operation and it is important to evaluate its performance for any other species considered for operation. 
The first run with PSI beams in the LHC was performed during MD studies in July 2018 \cite{Schaumann_2019_ipac_part_strip_ions,Schaumann:2670544}, where \leadpsi ions with one electron left were injected and stored in the LHC. While this experiment had as a main objective the demonstration of the possibility to store in the LHC PSI beams with good lifetime, it also gave the opportunity for the first tests of the PSI collimation process. 

The beam provided by the injectors for this experiment consisted of two \leadpsi bunches spaced by \unit[200]{ns} per SPS injection in the LHC. Each bunch featured an intensity of up to about $1.1\times10^{10}$ charges, or $1.3\times10^{8}$
ions. Because of machine protection requirements, the total circulating beam intensity had to stay below $3\times10^{11}$ charges during the  experiment. 
Table~\ref{tab:paramaters} lists all the beam and machine parameters used during the test. In Table~\ref{tab:summary} we list all detailed settings of the collimators used for the experiment and later for the simulation.

\begin{table}[h]
    \begin{center}
         \caption{Collection of beam and machine parameters used in experiments with \leadpsi and for the simulated beam in the SixTrack-FLUKA setup. The HL--LHC beam parameters are taken from Ref.~\cite{bruce20_HL_ion_report}.}
         \begin{tabular}{l  c c  }
      \hline 
      \hline 
      Parameter &   LHC MD & HL--LHC  \\
      \hline
      Ion   & $^{208}_{82}$Pb$^{81+}$ & $^{208}_{82}$Pb$^{81+}$ \\
      Equiv. proton beam energy   [TeV]& 6.5 & 7\\ 
      PSI beam energy   [TeV]& 526.5 \footnote{ it is \unit[6.5]{TeV}$\times Q$ where $Q=Z-1$ in this case is the charge number of the ion.}& 567 \\
      Proton beam emmitance  [$\mu$m]& 3.5 & 2.5 \\
      Ion beam emmitance  [$\mu$m] & 1.39 & 1.00 \\ 
      Bunch population   [Pb ions $\times10^{8}$]&  0.9 \footnote{first fill during the MD featured $1.3\times 10^8$ ions.}& 1.8 \\  
      Nb of bunches  & 6 & 1240 \\
      \hline
      $\beta^*$ in IP1/2/5/8 (Top energy)  [m]  & 1/1/1/1.5 & -- \\
      \hline 
      \hline
    \end{tabular}
    \label{tab:paramaters}
    \end{center}
\end{table}

The cleaning performance for \leadpsi was tested through dedicated loss maps at injection and at top energy. The latter case can be seen in Fig.~\ref{fig:example-loss-map} (bottom graph) together with the loss pattern obtained for \lead beams (top graph). Losses are normalized by the signal recorded at the primary collimator with highest losses. Both at injection and at top energy, severe losses are observed in the DS of IR7 with PSI beams. These losses turned out to be a real operational limitation, when a beam dump was triggered, two minutes after reaching top energy in the first fill. At the time, only 24 low-intensity bunches were stored, and the beam was dumped because of too high losses around $s=$\unit[20410]{m}, i.e. in the DS of IR7. In the second and last fill of the experiment, the number of bunches was reduced to six and the intensity per bunch was reduced to about $0.75\times10^{10}$ charges. This beam could successfully be accelerated to \unit[6.5]{Z TeV} and stored for about two hours~\cite{Schaumann_2019_ipac_part_strip_ions,Schaumann:2670544}. Still, the losses reached around 60\% of the dump threshold level.

\begin{table}[h]
\begin{center}
\caption{\label{tab:summary} Summary of the collimator settings at top energy optics as used in the 2018 MD, as well as for the HL--LHC collision optics. The settings are given in units of the nominal one sigma beam size, $\sigma _N$, expressed with respect to a proton-equivalent emittance of $\varepsilon=3.5\mu$m. }
  
  \begin{tabular}{ l r r }
  \hline 
  \hline
  Collimator name                 & LHC     &  HL--LHC  \\

                  & [$\sigma_N$]          & [$\sigma_N$]   \\ 
  \hline 
  \multicolumn{3}{l}{Betatron cleaning IR7}\\
  TCP  & 5.0 & 5.0  \\
  TCSG 
      & 6.5 & 6.5  \\
  TCLA & 10.0& 10.0  \\
  TCLD &n/a &14.0  \\
  
  \hline 
  \multicolumn{3}{l}{Momentum cleaning IR3}\\
  TCP  & 15.0 & 15.0 \\
  TCSG & 18.0 & 18.0  \\
  TCLA & 20.0& 20.0  \\
  \hline     
   \multicolumn{3}{l}{Experimental areas}\\
  TCT IR1/2/5 & 15 & 10  \\    
  TCT IR8  & 15    & 15 \\

  \hline 
  \hline
  \end{tabular} 

\end{center}
\end{table}

\subsection{Measured cleaning performance of PSI beams}

\begin{figure}[ht]
\centering	
\includegraphics[width=\columnwidth]{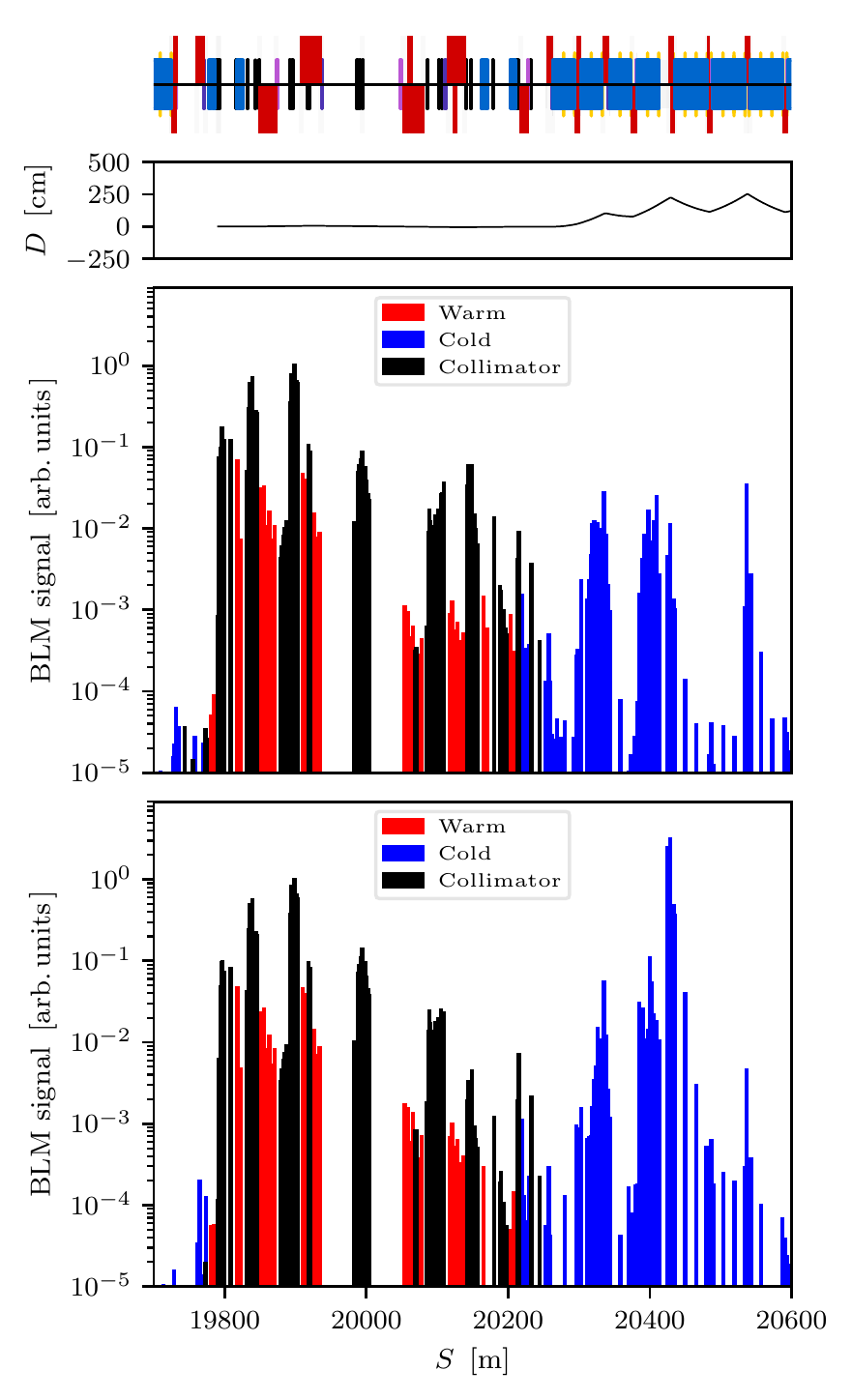}

\caption{Loss map for Beam~1 in the horizontal plane, recorded during the standard \lead fill (top) and during the experiments with \leadpsi beams that were carried out in LHC in 2018 at top energy (bottom). The PSI case shows a visible excess of the BLM signal around $s=$\unit[20430]{m}. The machine lattice and the horizontal locally generated dispersion from the TCP are shown on the very top of the figure.}
\label{fig:psi-lossmaps-b1h}
\end{figure}

In the second fill, the collimation performance at \mbox{\unit[6.5]{Z TeV}} was tested through loss maps. The worst losses were observed for Beam~1 in the horizontal plane (B1H)
and the measured loss map for this case can be seen in the bottom graph of  Fig.~\ref{fig:example-loss-map}, showing the full-ring loss pattern, and in the bottom graph of Fig.~\ref{fig:psi-lossmaps-b1h} (IR7 zoom). The peak losses were recorded by the BLMs located around the quadrupole magnet in cell 11,  at position $s\approx$\unit[20430]{m} from the center of IR1 (see for the local dispersion plotted in Fig.~\ref{fig:psi-lossmaps-b1h}). 
Please note, that Beam~2 did not reach the top energy, therefore there is no data for it \cite{Schaumann:2670544}.

The performance of the collimation system with PSI beams is significantly worse than for protons or fully stripped ion beams \cite{bruce14_PRSTAB_sixtr,hermes16_nim}. The recorded magnitude of the highest losses on the cold aperture of the DS, normalized to the intensity impacting on the TCP, is about 4~orders of magnitude larger than for standard proton operation, and about two orders of magnitude larger than for standard \lead operation, as can be seen in Fig.~\ref{fig:psi-lossmaps-b1h}. The BLM signal is also about 4~times larger at the peak in the DS than on the collimators, however, this does not mean that primary losses occur in the DS. Instead, this is due to the fact that the BLM response per locally lost particle is different at the two locations, because of the local geometry and materials. 


\section{Simulations of PSI beam losses in the LHC}
\label{sec:simulations}

\subsection{Process of DS losses through stripping action}

We hypothesize that the loss pattern in Fig.~\ref{fig:psi-lossmaps-b1h} can be explained by the stripping action of the collimators in combination with the increasing value of the  dispersion in the DS (see the top plot of the Fig.~\ref{fig:psi-lossmaps-b1h}). When passing through the TCP, partially stripped \leadpsi ions from the beam halo may lose their electron. In this stripping process, they do not experience an angular deflection sufficient to be intercepted by the TCSs. The resulting fully-stripped \lead ions have energies very close to nominal, i.e. \unit[7]{$Z$ TeV}, but have an altered charge-to-mass ratio and thus a  magnetic rigidity that differs from that of the circulating beam by about 1.2~\%. If they escape the betatron-cleaning insertion, the dispersion in the DS can push their trajectories onto the cold aperture.

The simulation tools used for this study do not support  the tracking of the partially stripped ions. Therefore, a simplified simulation setup was used, assuming  a 100\% stripping efficiency, i.e. PSI immediately losing their electron when they impact on a primary collimator such that any ions escaping the collimators are fully stripped. 
 
This assumption comes from the analysis of the mean free path (MFP) for the stripping process. The MFP was calculated to be \unit[0.04]{mm} for  \leadpsi ions in the Carbon-Fiber-Carbon material, of which the 0.6~m-long primary collimators are made of, using the methods in Ref.~\cite{Tolstikhina_2018}. 
The MFP can be compared to the distance traveled in the material for the different impact parameters, i.e. the distance between the collimator edge and the impact, where we assume that any impact takes place at the phase with maximum amplitude in phase space. This determines the impact angle. In this study we have considered impact parameter values \unit[0.1]{$\mu$m}, \unit[1]{$\mu$m} and  \unit[10]{$\mu$m} (as in Ref.~\cite{bruce14_PRSTAB_sixtr}) and we obtained traverse distances of: \unit[4.618]{mm}, \unit[46.18]{mm}, and \unit[461.8]{mm},  respectively. For each impact parameter value in  the given range, the distance traveled inside the collimator material is at least two orders of magnitude larger than the MFPs reported earlier. Therefore, it is a very good approximation to assume a full stripping of any PSI that approaches the TCPs. 


\subsection{Trajectories of fully stripped ions}

To test the theory of the stripping action of the collimators as explanation for the observed losses, we performed tracking simulations in MAD-X \cite{herr04,madx} of fully stripped ions emerging from the horizontal TCP. The trajectories of the fully stripped \lead ions were calculated with an effective $\Delta p/p=-1/82$ originating at one of the TCP jaws. Those trajectories were tracked through the betatron-cleaning insertion and the downstream DS, where the point at which they are intercepted by the aperture was calculated. 
In this simplified study, we do not assume any angular kicks at the TCP, but instead that the particles are at the phase of maximum horizontal excursion in phase space, as would be the case when they first hit the TCP after a slow diffusion process. 

\begin{figure}[pt!]
\centering
\includegraphics[width=\linewidth]{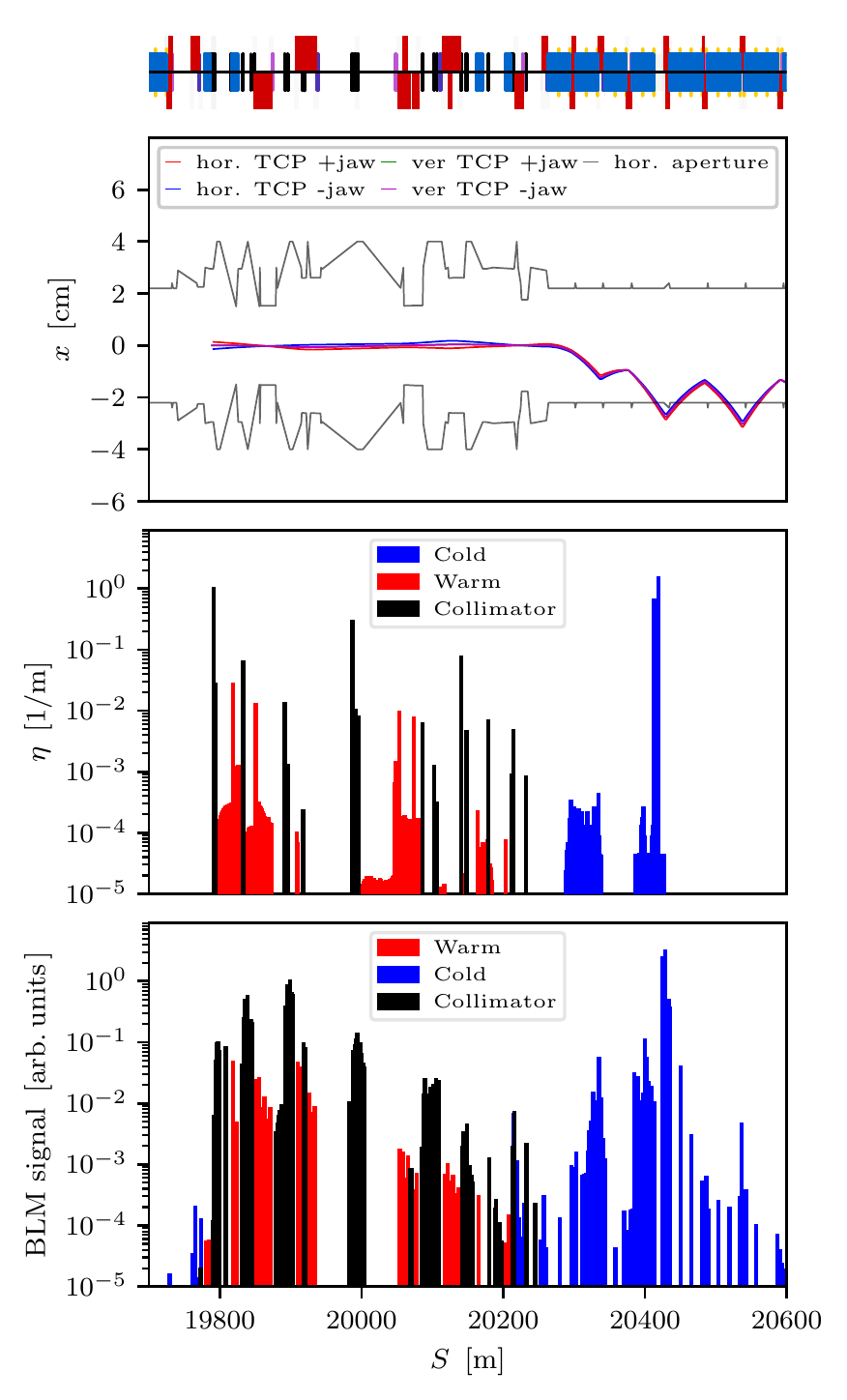}

\caption{The top plot shows trajectories of fully stripped off-rigidity \lead ions escaping the B1 TCPs, as calculated with MAD-X. The 4 trajectories depicted correspond to starting positions at both jaws of the horizontal and vertical TCPs. The middle plot shows the loss pattern simulated for PSI beams using the Sixtrack--FLUKA simulations. The bottom plot shows the measured  B1H loss map with \leadpsi beams at top energy. }
\label{fig:psi-lhc-experiment-fluka}
\end{figure}

A selected range of trajectories with the physical aperture overlaid is shown in Fig.~\ref{fig:psi-lhc-experiment-fluka}. As seen in the figure, the MAD-X trajectories of \lead escaping the TCP jaws bypass all downstream collimators and travel directly to the DS. We observe a calculated  loss position very similar to the one measured in the machine  (see Fig.~\ref{fig:psi-lossmaps-b1h}). This result strengthens the hypothesis on the origin of the large losses in the DS. Furthermore, it shows that since the loss mechanism involves dispersion, the loss location is relatively constant regardless of which TCP and jaw caused the stripping. 

\subsection{Simulations of the LHC experiment}

The next step to further investigate the stripping effect of the TCPs was to perform an integrated beam tracking and collimator interaction simulation. Using the coupling between the SixTrack \cite{schmidt94, sixtrack-web,bruce14_PRSTAB_sixtr} and FLUKA \cite{fluka14, Battistoni:2015epi} codes, described further in Refs.~\cite{mereghetti13_ipac,skordis18_tracking_workshop,pascal-thesis}, the aim was first to  reproduce the measurement results and second to validate the proposed mitigation solutions (see Section~\ref{sec:mitigations}). These simulations track beam halo particles through the magnetic lattice using SixTrack, accounting for a detailed aperture model to infer beam losses. When a particle hits a collimator, the particle-matter interaction is simulated using FLUKA, and any surviving particles that return to the beam vacuum are sent back to SixTrack for further magnetic tracking. Similar simulations have been extensively compared to measurements in previous publications for protons~\cite{bruce14_PRSTAB_sixtr,auchmann15_PRSTAB,bruce17_NIM_beta40cm,bruce19_PRAB_beam-halo_backgrounds_ATLAS} and Pb ions~\cite{hermes16_nim,hermes16_ipac_coupling}. 

The simulation was started at the upstream edge of the horizontal TCP,  tracking \lead ions with the electron already stripped, but in a machine configuration with the magnetic rigidity adjusted for \leadpsi, i.e. \unit[526.5]{TeV}.  To simulate the experiment we used $5\times10^5$ macro particles hitting a collimator with an impact parameter of \unit[1]{$\mu$m} (to remain conform to the regular \lead simulations). The cleaning hierarchy setup was reproduced as in the experiment, and it is listed in detail in Table~\ref{tab:summary}.

Figure~\ref{fig:psi-lhc-experiment-fluka} shows the simulated loss distribution for the LHC configuration used during the PSI machine test, including the same  optics and collimator settings. One can see the very good qualitative agreement between the measured loss map and the simulated one, with the highest cold loss peak at the place of the aperture impact predicted with MAD-X \cite{madx}, used to estimate the single-pass trajectory for off-momentum particles. While the agreement on the peak losses (see Fig.~\ref{fig:psi-lhc-experiment-fluka}) is visible in the DS region around $s=\unit[20400]{m}$, some differences may be noticed in the TCP/TCS region ($s=\unit[19800-19900]{m}$). This apparent discrepancy could be explained by the fact that a full quantitative comparison in Fig.~\ref{fig:psi-lhc-experiment-fluka} cannot be made, since the BLM measurement is sensitive to the secondary shower particles that emerge outside of the impacted elements, while our simulations show the number of primary nuclei impacting on the aperture or disintegrating on the collimators. The measured loss pattern is also affected by a cross-talk between nearby BLMs, where any BLM intercepts the shower not only from the element at which it is placed, but also from losses on nearby upstream elements. Previous studies for protons have shown that the agreement improves dramatically when a further simulation of the shower development and the energy deposition in the BLM is performed~\cite{bruce14_PRSTAB_sixtr, auchmann15_PRSTAB}. 

\begin{figure}[t]
\centering
 \includegraphics[width=\columnwidth]{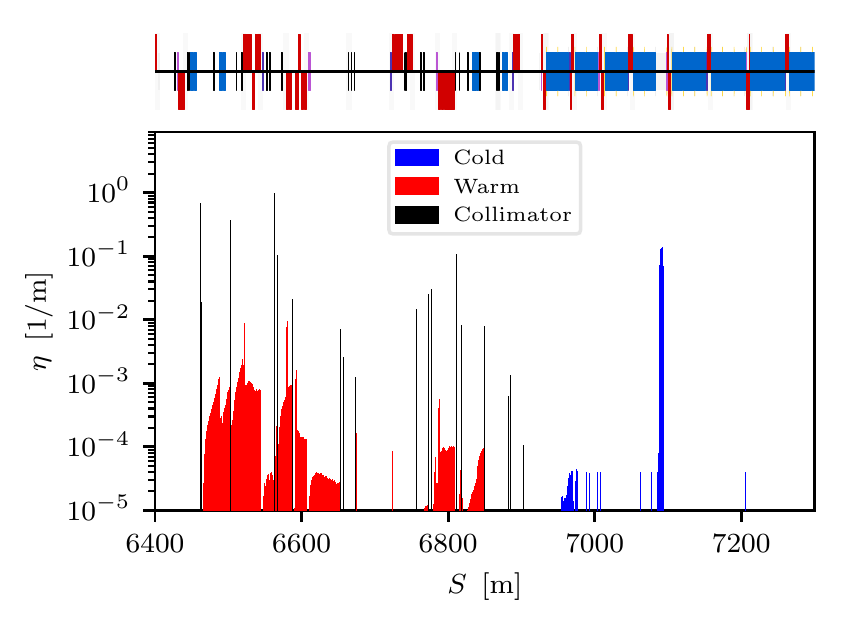}
\caption{SixTrack--FLUKA simulation results for Beam~2. The presence of an excess losses in the first cell after the dispersion suppressor (around $s=$\unit[7100]{m}) is visible, as for the case of Beam~1 (see Fig.\ref{fig:psi-lhc-experiment-fluka}).}
\label{fig:psi-fluka-lhc-beam2}
\end{figure}

Although the experiment was only performed for Beam~1, a simulation for Beam~2 was also carried out. A similar loss peak as for Beam~1 was noticed in the simulation results. However, due to small asymmetries of the optics functions with respect to Beam~1, the Beam~2 loss peak is found more downstream. Figure~\ref{fig:psi-fluka-lhc-beam2} shows the result of the simulation for Beam~2. 

The main loss position of fully stripped Pb ions given by the trajectories and tracking simulations in Fig.~\ref{fig:psi-lhc-experiment-fluka} is a strong indirect demonstration that the stripping process is the main source of the observed worsening in cleaning performance.  

\section{Mitigation techniques}
\label{sec:mitigations}

\subsection{Dispersion suppressor collimators}

The High Luminosity LHC (HL--LHC)~\cite{hl-lhc-tech-design}  will start operating in 2027 and will push forward the luminosity frontier, increasing by about a factor of 10 the LHC integrated luminosity. Several hardware upgrades will be implemented for the collimation system of the HL--LHC. Among other upgrades not relevant for this work, new dispersion suppressor collimators -- called TCLD --  will be installed to protect the downstream DS region~\cite{redaelli14_chamonix}.  During the current long shutdown period (LS2, in 2019-2021) it is planned to install one TCLD collimator per beam in the DS regions around IR7. The primary purpose of those collimators is to intercept dispersive losses coming from the betatron-collimation insertion and studies show that their presence can reduce the losses in the DS for both proton and ion beams \cite{hl-lhc-tech-design,bruce14ipac_DS_coll,hermes15_ipac}. To make space for the TCLD, a standard 8.33\,$\mathrm{T}$, 15~m-long main dipole will be replaced by two shorter, 5~m-long 11\,$\mathrm{T}$ dipoles, based on the Nb$_3$Sn superconducting alloy, with the collimator assembly in the middle \cite{Zlobin:2019ven}. The planned location for the TCLD installation at the longitudinal position $s$=\unit[20310]{m}, see Fig.6. 

\begin{figure}[t]
\centering
     \includegraphics[width=\linewidth]{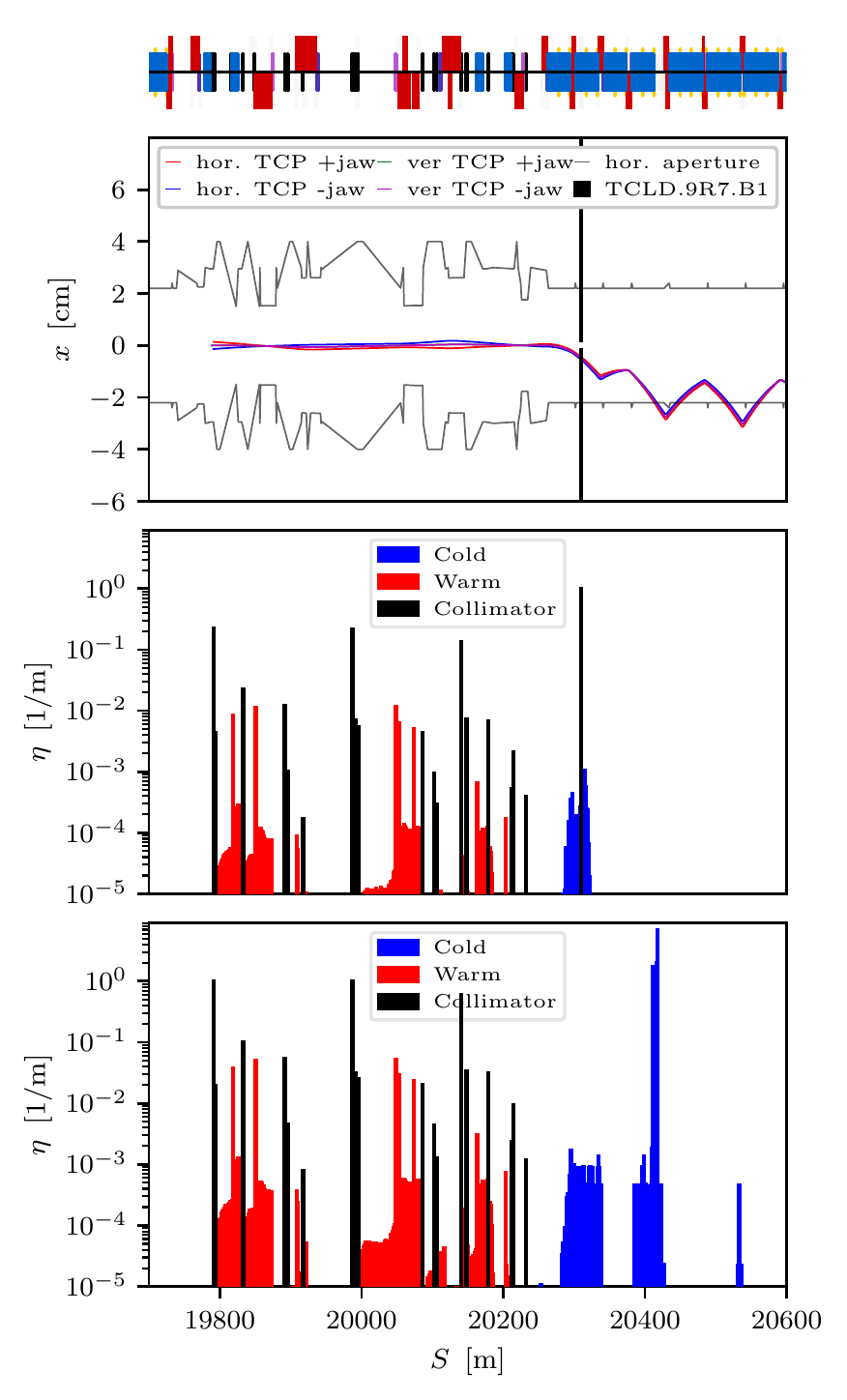}
\caption{\label{fig:hllhc-tcld-in-trajectories-and-losses}The top figure shows trajectories calculated with MAD-X of fully stripped off-rigidity \lead ions escaping the TCPs together with the aperture model and the TCLD with of 14\,$\sigma$ opening. The middle plot shows the SixTrack-FLUKA coupling simulation of the loss map for HL--LHC version 1.2, including the TCLD at 14~$\sigma$ in cell 9. The bottom plot shows the simulated loss map for the same optics but without the TCLD, resulting in the similar loss pattern as observed in the LHC.}
\end{figure}

Since the purpose of the TCLD is to catch dispersive losses, we investigate in detail whether it could  potentially be used to intercept the fully stripped \lead ions during PSI operation. A plot of the \lead trajectories after electron stripping at the TCPs is shown in the top graph of Fig.~\ref{fig:hllhc-tcld-in-trajectories-and-losses}. The longitudinal position of the TCLD is indicated by the black line, showing also its expected operational opening of 14\,$\sigma$. The aperture of the fixed layout elements is also shown. It is shown that the TCLD can indeed intercept the fully-stripped \lead ions before they reach the cold aperture of DS magnets. 

Complete loss maps simulations were performed to confirm more quantitatively this finding. The collimator settings of Table~\ref{tab:summary} were considered. The simulated loss distribution is shown in the middle and bottom graphs of Fig.~\ref{fig:hllhc-tcld-in-trajectories-and-losses}, for the cases with and without TCLD, respectively. The TCLD efficiently catches the fully stripped ions, reducing the downstream cold losses by about four orders of magnitude. It is also noticeable that the TCLD collimator becomes (in case of operation with \leadpsi) the collimator with the highest fraction of the energy absorbed. 

Another option including the dispersion suppressor collimator, would be to install one additional unit in the empty cryostat. The added values oh that solution would be lack of need for additional \unit[11]{T} magnets, and absorber's closer position to the peak losses. Moreover orchestrating the setting in a way to have the losses split between the two collimators is potentially possible. But detailed look in this aspect goes beyond the scope of this paper. 

\subsection{Power load on \unit[11]{T} coil and TCLD collimator}

Even though the TCLD effectively intercepts the losses of fully stripped ions, the risk of quenching the downstream magnets must be assessed by taking into account the energy leaking out of the TCLD. To estimate this risk, we assume a quench limit of \unit[70]{mW/cm$^3$} for the \unit[11]{T} dipole~\cite{bottura18}, and a \unit[0.2]{h} minimum beam lifetime, according to the design specification of the HL--LHC collimation system. We also assume the maximum intensity for the \leadpsi beams as considered for the \lead for HL--LHC, namely $2.2\times10^{11}$~\cite{bruce20_HL_ion_report, hl-lhc-tech-design}.  These assumptions correspond to a loss rate of about $3.1\times10^8$ ions per second on the TCP, carrying a total power of about 28~kW~\cite{hl-lhc-tech-design}. Furthermore, we assume that each \lead ion impacting on the TCLD causes an energy deposition of \unit[$5\times10^{-7}$]{mJ/cm$^3$}  in the coils of the downstream magnet \cite{anton-et-all-on-TCLD-losses}. This number is extracted from an energy deposition study with  FLUKA of betatron losses during standard \lead operation. However, it should be noted that this is a very approximate number, since the impact distribution on the TCLD  will be different for PSI losses. Therefore, for a more detailed assessment, specific energy deposition simulations should be repeated for this case. The total maximum \leadpsi beam intensity that is acceptable without quenching can then be calculated to $3\times 10^{11}$~Pb ions, which is beyond the baseline Pb total intensity for the ion runs in HL--LHC.  Therefore, it is not expected that the total PSI intensity will be severely limited by the shower on the downstream 11~T magnet. To refine the intensity estimate, energy deposition simulations should be performed using the realistic PSI impact distribution on the TCLD.

From the Fig.~\ref{fig:hllhc-tcld-in-trajectories-and-losses} the power load on the TCLD during a 0.2~h beam lifetime drop is estimated to about 30~\% of the total beam losses, i.e. about \unit[9]{kW}. A simple scaling of previous studies on other tungsten collimators~\cite{TCT_fede} shows that the TCLD risks to have a peak temperature of 150~$^\circ$C or more, and that it could temporarily deform by a few hundred microns but without any permanent damage to the collimator. Therefore, we do not expect severe limitations on the total beam intensity below the assumed baseline. However, a more detailed assessment, including further simulations of the energy deposition and thermo-mechanical response of the TCLD, is needed to draw a firm conclusion. In addition to the 0.2~h beam lifetime scenario, the impact of steady-state losses with a 1~h beam lifetime could also be studied as done for the standard HL--LHC running scenario~\cite{hl-lhc-tech-design}.  


Since the TCLDs are planned to be installed in LS2, the LHC run~3 starting in 2021 will provide a unique opportunity to perform dedicated tests of cleaning efficiency with \leadpsi ion beams with the final DS layouts.

\subsection{Alternative mitigation measures}
In case of problems with the TCLD collimators, or if a further mitigation would be needed, other alleviation techniques could be envisaged. We discuss these concepts here, although without detailed simulations. 

Crystal collimation is a novel technique being investigated for the LHC and HL--LHC \cite{daniele-thesis,scandale16,Mirarchi2017_crystals}. A bent silicon crystal is used
instead of the amorphous carbon TCP in the standard collimation setup. Beam halo particles incident on the crystal can enter a channeling regime, in which their trajectories are guided by the potential between crystalline planes. The angular deflection achieved by channeling in the crystal is much larger than the deflection achieved by scattering in an amorphous material and beam halo particles can be directed onto a single massive absorber with a large impact parameter, reducing the need for additional secondary collimators and additional absorbers. 

While the interaction of the PSI with the crystal is currently not well characterised, crystal collimation has shown promise for improving the cleaning efficiency with heavy ion beams~\cite{D'Andrea:2678781} and it is proposed to study if it can also alleviate the losses for PSI beams.  
If the impacting ions could be channelled by the crystal even if their electron has been stripped, they could could be steered onto the absorber as in the standard crystal collimation scheme. This concept requires thorough experimental feasibility studies before being relied upon. 
A crystal test stand installation is available in the LHC for collimation studies \cite{Mirarchi2017_crystals}, and it is planned to test this with PSI beams when available again in the LHC. 


As the stripping action of collimators produces a secondary beam of similar particles, another mitigation strategy may involve an orbit bump to shift the losses to a less sensitive location. Orbit bumps have been successfully employed for the case of secondary beams created in the collisions between \lead beams, formed by the process of Bound-Free Pair Production (BFPP)~\cite{klein01,prl07,prstabBFPP09,schaumann16_md_BFPPquench,jowett16_ipac_bfpp}. The bump is used to shift the impact location of the secondary beam on the mechanical aperture out of the main dipole and into an empty connection cryostat so that no impacts occur directly on superconducting magnets. 

A similar strategy can be considered for PSI secondary beams. A local closed orbit bump could be used to optimize the loss location of the fully stripped \lead beam, presently around $s$=\unit[20400]{m}, to the nearby empty connection cryostat. However, a rather large bump amplitude of about 6.5~mm would be needed to move the losses away from the quadrupole, and it needs to be studied if such a bump is operationally feasible.  It should also be noted that the total peak power of the local losses on the connection cryostat during PSI operation will be much higher than the power of the BFPP losses that were successfully handled in the LHC. 

The peak collimation losses are assumed to be transient and last only for a few seconds, but steady collimation losses during standard operation might also cause limitations. Therefore, to conclude on the feasibility of this mitigation strategy, the power deposition in the connection cryostat and downstream magnets needs to be studied in further detail for different scenarios of losses on the TCPs.


\section{Conclusions and outlook}

Partially stripped \leadpsi ions with one electron left were injected, accelerated, and stored in the LHC  for the first time. The results of the first measurement of the collimation performance with those beams at the LHC show that the cleaning efficiency is prohibitively low for high-intensity operation, due to very high localized losses in the dispersion suppressor. We have presented  studies showing that the likely reason for the poor collimation performance is the stripping action of the primary collimators, which causes nearly every ion that touches the collimator to lose its electron. If an ion does not fragment in the collimator, it re-enters the beam with a higher charge, causing it to be lost in the dispersion suppressor where the dispersion rises. 

For a detailed investigation of the observed losses, tracking simulations were performed. The simulated impact location of the fully stripped \lead ions, as well as the finding that this loss location is by far dominating, are in excellent agreement with the measurements. This demonstrates that the stripping action of the material in the primary collimators is indeed very likely responsible for the observed loss peaks, which have never been observed before at this magnitude with other particle species. 

The simulations were extended for the machine layout including  upgrades foreseen for HL--LHC. Several mitigation strategies are under consideration for reducing the losses on the dispersion suppressor and thus increasing the intensity reach of partially stripped ions operation. The use of a new TCLD collimator, which is scheduled to be installed before the next LHC run, was identified as the most promising option, as it could efficiently intercept the fully stripped \lead ions before reaching the cold magnets. A preliminary estimate extrapolated from energy deposition studies for \leadpsi indicates that the magnet downstream of the TCLD is not likely to quench with the assumption of beam lifetime and total beam intensity as for the \lead beam, even with a potential full HL--LHC PSI beam. On the other hand, there is a risk that the TCLD itself could suffer significant deformations, potentially limiting the maximum intensity to a factor of a few below the nominal HL--LHC design intensity for \lead. 

However, additional comprehensive energy deposition studies should be carried out to better quantify the current beam intensity limit of safe operation for both the magnets and the TCLD. The non-standard beam particle type and the additional physical interactions the PSI can undergo are not supported by the available simulation tools. An active effort is directed to extending the existing simulation frameworks and developing new ones that would enable future studies of various PSI collimation aspects.

Two other mitigation techniques were discussed: crystal collimation and dedicated orbit bumps. Both options, alone or in combination with other mitigation techniques, may be considered as useful in case the TCLD would turn out not being sufficient regarding the quench limits of the nearby magnets, or if the load on the TCLD itself would be too high. 

\begin{acknowledgments}
The authors would like to thank the LHC operations crew for their help during the measurements. Additionally, the authors thank  F.~Carra for the detailed and instructive comments on the tungsten collimator power-deposition loads.
\end{acknowledgments}


\bibliography{references}

\begin{thebibliography}{55}%
\makeatletter
\providecommand \@ifxundefined [1]{%
 \@ifx{#1\undefined}
}%
\providecommand \@ifnum [1]{%
 \ifnum #1\expandafter \@firstoftwo
 \else \expandafter \@secondoftwo
 \fi
}%
\providecommand \@ifx [1]{%
 \ifx #1\expandafter \@firstoftwo
 \else \expandafter \@secondoftwo
 \fi
}%
\providecommand \natexlab [1]{#1}%
\providecommand \enquote  [1]{``#1''}%
\providecommand \bibnamefont  [1]{#1}%
\providecommand \bibfnamefont [1]{#1}%
\providecommand \citenamefont [1]{#1}%
\providecommand \href@noop [0]{\@secondoftwo}%
\providecommand \href [0]{\begingroup \@sanitize@url \@href}%
\providecommand \@href[1]{\@@startlink{#1}\@@href}%
\providecommand \@@href[1]{\endgroup#1\@@endlink}%
\providecommand \@sanitize@url [0]{\catcode `\\12\catcode `\$12\catcode
  `\&12\catcode `\#12\catcode `\^12\catcode `\_12\catcode `\%12\relax}%
\providecommand \@@startlink[1]{}%
\providecommand \@@endlink[0]{}%
\providecommand \url  [0]{\begingroup\@sanitize@url \@url }%
\providecommand \@url [1]{\endgroup\@href {#1}{\urlprefix }}%
\providecommand \urlprefix  [0]{URL }%
\providecommand \Eprint [0]{\href }%
\providecommand \doibase [0]{https://doi.org/}%
\providecommand \selectlanguage [0]{\@gobble}%
\providecommand \bibinfo  [0]{\@secondoftwo}%
\providecommand \bibfield  [0]{\@secondoftwo}%
\providecommand \translation [1]{[#1]}%
\providecommand \BibitemOpen [0]{}%
\providecommand \bibitemStop [0]{}%
\providecommand \bibitemNoStop [0]{.\EOS\space}%
\providecommand \EOS [0]{\spacefactor3000\relax}%
\providecommand \BibitemShut  [1]{\csname bibitem#1\endcsname}%
\let\auto@bib@innerbib\@empty
\bibitem [{\citenamefont {Br{\"{u}}ning}\ \emph {et~al.}(2004)\citenamefont
  {Br{\"{u}}ning}, \citenamefont {Collier}, \citenamefont {Lebrun},
  \citenamefont {Myers}, \citenamefont {Ostojic}, \citenamefont {Poole},\ and\
  \citenamefont {{P.~Proudlock (editors)}}}]{lhcdesignV1}%
  \BibitemOpen
  \bibfield  {author} {\bibinfo {author} {\bibfnamefont {O.~S.}\ \bibnamefont
  {Br{\"{u}}ning}}, \bibinfo {author} {\bibfnamefont {P.}~\bibnamefont
  {Collier}}, \bibinfo {author} {\bibfnamefont {P.}~\bibnamefont {Lebrun}},
  \bibinfo {author} {\bibfnamefont {S.}~\bibnamefont {Myers}}, \bibinfo
  {author} {\bibfnamefont {R.}~\bibnamefont {Ostojic}}, \bibinfo {author}
  {\bibfnamefont {J.}~\bibnamefont {Poole}},\ and\ \bibinfo {author}
  {\bibnamefont {{P.~Proudlock (editors)}}},\ }\bibfield  {title} {\bibinfo
  {title} {{LHC} design report v.1 : The {LHC} main ring},\ }\href@noop {}
  {\bibfield  {journal} {\bibinfo  {journal} {CERN-2004-003-V1}\ } (\bibinfo
  {year} {2004})}\BibitemShut {NoStop}%
\bibitem [{\citenamefont {Citron}\ \emph {et~al.}(2018)\citenamefont {Citron}
  \emph {et~al.}}]{YR_WG5_2018}%
  \BibitemOpen
  \bibfield  {author} {\bibinfo {author} {\bibfnamefont {Z.}~\bibnamefont
  {Citron}} \emph {et~al.},\ }\bibfield  {title} {\bibinfo {title} {{Future
  physics opportunities for high-density QCD at the LHC with heavy-ion and
  proton beams}},\ }\bibfield  {booktitle} {\emph {\bibinfo {booktitle}
  {{HL/HE-LHC Workshop: Workshop on the Physics of HL-LHC, and Perspectives at
  HE-LHC Geneva, Switzerland, June 18-20, 2018}}},\ }\href
  {https://cds.cern.ch/record/2650176?ln=en} {\bibfield  {journal} {\bibinfo
  {journal} {CERN-LPCC-2018-07}\ } (\bibinfo {year} {2018})},\ \Eprint
  {https://arxiv.org/abs/1812.06772} {arXiv:1812.06772 [hep-ph]} \BibitemShut
  {NoStop}%
\bibitem [{\citenamefont {Schaumann}\ \emph {et~al.}(2018)\citenamefont
  {Schaumann} \emph {et~al.}}]{Schaumann:IPAC2018-MOPMF039}%
  \BibitemOpen
  \bibfield  {author} {\bibinfo {author} {\bibfnamefont {M.}~\bibnamefont
  {Schaumann}} \emph {et~al.},\ }\bibfield  {title} {\bibinfo {title} {{F}irst
  {X}enon{-X}enon {C}ollisions in the {LHC}},\ }in\ \href
  {https://doi.org/doi:10.18429/JACoW-IPAC2018-MOPMF039} {\emph {\bibinfo
  {booktitle} {Proc. 9th International Particle Accelerator Conference
  (IPAC'18), Vancouver, BC, Canada, 29 April-04 May 2018}}},\ \bibinfo {series
  and number} {\bibinfo {series} {International Particle Accelerator
  Conference}\ No.~\bibinfo {number} {9}}\ (\bibinfo  {publisher} {JACoW
  Publishing},\ \bibinfo {address} {Geneva, Switzerland},\ \bibinfo {year}
  {2018})\ pp.\ \bibinfo {pages} {180--183},\ \bibinfo {note}
  {https://doi.org/10.18429/JACoW-IPAC2018-MOPMF039}\BibitemShut {NoStop}%
\bibitem [{\citenamefont {Alemany}\ \emph {et~al.}(2019)\citenamefont
  {Alemany}, \citenamefont {Burrage}, \citenamefont {Bartosik}, \citenamefont
  {Bernhard}, \citenamefont {Boyd}, \citenamefont {Brugger}, \citenamefont
  {Calviani}, \citenamefont {Carli}, \citenamefont {Charitonidis},
  \citenamefont {Curtin}, \citenamefont {Dainese}, \citenamefont {de~Roeck},
  \citenamefont {Diehl}, \citenamefont {Döbrich}, \citenamefont {Evans},
  \citenamefont {Feng}, \citenamefont {Ferro-Luzzi}, \citenamefont {Gatignon},
  \citenamefont {Gilardoni}, \citenamefont {Gninenko}, \citenamefont
  {Graziani}, \citenamefont {Gschwendtner}, \citenamefont {Goddard},
  \citenamefont {Hartin}, \citenamefont {Irastorza}, \citenamefont {Jaeckel},
  \citenamefont {Jacobsson}, \citenamefont {Jungmann}, \citenamefont {Kirch},
  \citenamefont {Kling}, \citenamefont {Krasny}, \citenamefont {Lamont},
  \citenamefont {Lanfranchi}, \citenamefont {Lansberg}, \citenamefont
  {Lindner}, \citenamefont {Long}, \citenamefont {Magnon}, \citenamefont
  {Mallot}, \citenamefont {Vidal}, \citenamefont {Moulson}, \citenamefont
  {Papucci}, \citenamefont {Pawlowski}, \citenamefont {Pedraza}, \citenamefont
  {Petridis}, \citenamefont {Pospelov}, \citenamefont {Pulawski}, \citenamefont
  {Redaelli}, \citenamefont {Rozanov}, \citenamefont {Rumolo}, \citenamefont
  {Ruoso}, \citenamefont {Schacher}, \citenamefont {Schnell}, \citenamefont
  {Schuster}, \citenamefont {Semertzidis}, \citenamefont {Siemko},
  \citenamefont {Spadaro}, \citenamefont {Stapnes}, \citenamefont {Stocchi},
  \citenamefont {Ströher}, \citenamefont {Usai}, \citenamefont {Vallée},
  \citenamefont {Venanzoni}, \citenamefont {Wilkinson},\ and\ \citenamefont
  {Wing}}]{alemany2019summary}%
  \BibitemOpen
  \bibfield  {author} {\bibinfo {author} {\bibfnamefont {R.}~\bibnamefont
  {Alemany}}, \bibinfo {author} {\bibfnamefont {C.}~\bibnamefont {Burrage}},
  \bibinfo {author} {\bibfnamefont {H.}~\bibnamefont {Bartosik}}, \bibinfo
  {author} {\bibfnamefont {J.}~\bibnamefont {Bernhard}}, \bibinfo {author}
  {\bibfnamefont {J.}~\bibnamefont {Boyd}}, \bibinfo {author} {\bibfnamefont
  {M.}~\bibnamefont {Brugger}}, \bibinfo {author} {\bibfnamefont
  {M.}~\bibnamefont {Calviani}}, \bibinfo {author} {\bibfnamefont
  {C.}~\bibnamefont {Carli}}, \bibinfo {author} {\bibfnamefont
  {N.}~\bibnamefont {Charitonidis}}, \bibinfo {author} {\bibfnamefont
  {D.}~\bibnamefont {Curtin}}, \bibinfo {author} {\bibfnamefont
  {A.}~\bibnamefont {Dainese}}, \bibinfo {author} {\bibfnamefont
  {A.}~\bibnamefont {de~Roeck}}, \bibinfo {author} {\bibfnamefont
  {M.}~\bibnamefont {Diehl}}, \bibinfo {author} {\bibfnamefont
  {B.}~\bibnamefont {Döbrich}}, \bibinfo {author} {\bibfnamefont
  {L.}~\bibnamefont {Evans}}, \bibinfo {author} {\bibfnamefont {J.~L.}\
  \bibnamefont {Feng}}, \bibinfo {author} {\bibfnamefont {M.}~\bibnamefont
  {Ferro-Luzzi}}, \bibinfo {author} {\bibfnamefont {L.}~\bibnamefont
  {Gatignon}}, \bibinfo {author} {\bibfnamefont {S.}~\bibnamefont {Gilardoni}},
  \bibinfo {author} {\bibfnamefont {S.}~\bibnamefont {Gninenko}}, \bibinfo
  {author} {\bibfnamefont {G.}~\bibnamefont {Graziani}}, \bibinfo {author}
  {\bibfnamefont {E.}~\bibnamefont {Gschwendtner}}, \bibinfo {author}
  {\bibfnamefont {B.}~\bibnamefont {Goddard}}, \bibinfo {author} {\bibfnamefont
  {A.}~\bibnamefont {Hartin}}, \bibinfo {author} {\bibfnamefont
  {I.}~\bibnamefont {Irastorza}}, \bibinfo {author} {\bibfnamefont
  {J.}~\bibnamefont {Jaeckel}}, \bibinfo {author} {\bibfnamefont
  {R.}~\bibnamefont {Jacobsson}}, \bibinfo {author} {\bibfnamefont
  {K.}~\bibnamefont {Jungmann}}, \bibinfo {author} {\bibfnamefont
  {K.}~\bibnamefont {Kirch}}, \bibinfo {author} {\bibfnamefont
  {F.}~\bibnamefont {Kling}}, \bibinfo {author} {\bibfnamefont
  {W.}~\bibnamefont {Krasny}}, \bibinfo {author} {\bibfnamefont
  {M.}~\bibnamefont {Lamont}}, \bibinfo {author} {\bibfnamefont
  {G.}~\bibnamefont {Lanfranchi}}, \bibinfo {author} {\bibfnamefont {J.-P.}\
  \bibnamefont {Lansberg}}, \bibinfo {author} {\bibfnamefont {A.}~\bibnamefont
  {Lindner}}, \bibinfo {author} {\bibfnamefont {K.}~\bibnamefont {Long}},
  \bibinfo {author} {\bibfnamefont {A.}~\bibnamefont {Magnon}}, \bibinfo
  {author} {\bibfnamefont {G.}~\bibnamefont {Mallot}}, \bibinfo {author}
  {\bibfnamefont {F.~M.}\ \bibnamefont {Vidal}}, \bibinfo {author}
  {\bibfnamefont {M.}~\bibnamefont {Moulson}}, \bibinfo {author} {\bibfnamefont
  {M.}~\bibnamefont {Papucci}}, \bibinfo {author} {\bibfnamefont {J.~M.}\
  \bibnamefont {Pawlowski}}, \bibinfo {author} {\bibfnamefont {I.}~\bibnamefont
  {Pedraza}}, \bibinfo {author} {\bibfnamefont {K.}~\bibnamefont {Petridis}},
  \bibinfo {author} {\bibfnamefont {M.}~\bibnamefont {Pospelov}}, \bibinfo
  {author} {\bibfnamefont {S.}~\bibnamefont {Pulawski}}, \bibinfo {author}
  {\bibfnamefont {S.}~\bibnamefont {Redaelli}}, \bibinfo {author}
  {\bibfnamefont {S.}~\bibnamefont {Rozanov}}, \bibinfo {author} {\bibfnamefont
  {G.}~\bibnamefont {Rumolo}}, \bibinfo {author} {\bibfnamefont
  {G.}~\bibnamefont {Ruoso}}, \bibinfo {author} {\bibfnamefont
  {J.}~\bibnamefont {Schacher}}, \bibinfo {author} {\bibfnamefont
  {G.}~\bibnamefont {Schnell}}, \bibinfo {author} {\bibfnamefont
  {P.}~\bibnamefont {Schuster}}, \bibinfo {author} {\bibfnamefont
  {Y.}~\bibnamefont {Semertzidis}}, \bibinfo {author} {\bibfnamefont
  {A.}~\bibnamefont {Siemko}}, \bibinfo {author} {\bibfnamefont
  {T.}~\bibnamefont {Spadaro}}, \bibinfo {author} {\bibfnamefont
  {S.}~\bibnamefont {Stapnes}}, \bibinfo {author} {\bibfnamefont
  {A.}~\bibnamefont {Stocchi}}, \bibinfo {author} {\bibfnamefont
  {H.}~\bibnamefont {Ströher}}, \bibinfo {author} {\bibfnamefont
  {G.}~\bibnamefont {Usai}}, \bibinfo {author} {\bibfnamefont {C.}~\bibnamefont
  {Vallée}}, \bibinfo {author} {\bibfnamefont {G.}~\bibnamefont {Venanzoni}},
  \bibinfo {author} {\bibfnamefont {G.}~\bibnamefont {Wilkinson}},\ and\
  \bibinfo {author} {\bibfnamefont {M.}~\bibnamefont {Wing}},\ }\href@noop {}
  {\bibinfo {title} {Summary report of physics beyond colliders at cern}}
  (\bibinfo {year} {2019}),\ \Eprint {https://arxiv.org/abs/1902.00260}
  {arXiv:1902.00260 [hep-ex]} \BibitemShut {NoStop}%
\bibitem [{\citenamefont {Krasny}(2015)}]{krasny2015gamma}%
  \BibitemOpen
  \bibfield  {author} {\bibinfo {author} {\bibfnamefont {M.~W.}\ \bibnamefont
  {Krasny}},\ }\href@noop {} {\bibinfo {title} {The gamma factory proposal for
  cern}} (\bibinfo {year} {2015}),\ \Eprint {https://arxiv.org/abs/1511.07794}
  {arXiv:1511.07794 [hep-ex]} \BibitemShut {NoStop}%
\bibitem [{\citenamefont {Krasny}\ \emph {et~al.}(2019)\citenamefont {Krasny},
  \citenamefont {Abramov}, \citenamefont {Alden}, \citenamefont
  {Alemany~Fernandez}, \citenamefont {Antsiferov}, \citenamefont {Apyan},
  \citenamefont {Bartosik}, \citenamefont {Bessonov}, \citenamefont
  {Biancacci}, \citenamefont {Bieron}, \citenamefont {Bogacz}, \citenamefont
  {Bosco}, \citenamefont {Bruce}, \citenamefont {Budker}, \citenamefont
  {Cassou}, \citenamefont {Castelli}, \citenamefont {Chaikovska}, \citenamefont
  {Curatolo}, \citenamefont {Czodrowski}, \citenamefont {Derevianko},
  \citenamefont {Dupraz}, \citenamefont {Dutheil}, \citenamefont {Dzierzega},
  \citenamefont {Fedosseev}, \citenamefont {Fuster~Martinez}, \citenamefont
  {Gibson}, \citenamefont {Goddard}, \citenamefont {Gorzawski}, \citenamefont
  {Hirlander}, \citenamefont {Jowett}, \citenamefont {Kersevan}, \citenamefont
  {Kowalska}, \citenamefont {Kroeger}, \citenamefont {Kuchler}, \citenamefont
  {Lamont}, \citenamefont {Lefevre}, \citenamefont {Manglunki}, \citenamefont
  {Marsh}, \citenamefont {Martens}, \citenamefont {Molson}, \citenamefont
  {Nutarelli}, \citenamefont {Nevay}, \citenamefont {Petrenko}, \citenamefont
  {Petrillo}, \citenamefont {Placzek}, \citenamefont {Redaelli}, \citenamefont
  {Peinaud}, \citenamefont {Pustelny}, \citenamefont {Rochester}, \citenamefont
  {Sapinski}, \citenamefont {Schaumann}, \citenamefont {Scrivens},
  \citenamefont {Serafini}, \citenamefont {Shevelko}, \citenamefont
  {Stoehlker}, \citenamefont {Surzhykov}, \citenamefont {Tolstikhina},
  \citenamefont {Velotti}, \citenamefont {Weber}, \citenamefont {Wu},
  \citenamefont {Yin-Vallgren}, \citenamefont {Zanetti}, \citenamefont
  {Zimmermann}, \citenamefont {Zolotorev},\ and\ \citenamefont
  {Zomer}}]{Krasny:2690736}%
  \BibitemOpen
  \bibfield  {author} {\bibinfo {author} {\bibfnamefont {M.~W.}\ \bibnamefont
  {Krasny}}, \bibinfo {author} {\bibfnamefont {A.}~\bibnamefont {Abramov}},
  \bibinfo {author} {\bibfnamefont {S.~E.}\ \bibnamefont {Alden}}, \bibinfo
  {author} {\bibfnamefont {R.}~\bibnamefont {Alemany~Fernandez}}, \bibinfo
  {author} {\bibfnamefont {P.~S.}\ \bibnamefont {Antsiferov}}, \bibinfo
  {author} {\bibfnamefont {A.}~\bibnamefont {Apyan}}, \bibinfo {author}
  {\bibfnamefont {H.}~\bibnamefont {Bartosik}}, \bibinfo {author}
  {\bibfnamefont {E.~G.}\ \bibnamefont {Bessonov}}, \bibinfo {author}
  {\bibfnamefont {N.}~\bibnamefont {Biancacci}}, \bibinfo {author}
  {\bibfnamefont {J.}~\bibnamefont {Bieron}}, \bibinfo {author} {\bibfnamefont
  {A.}~\bibnamefont {Bogacz}}, \bibinfo {author} {\bibfnamefont
  {A.}~\bibnamefont {Bosco}}, \bibinfo {author} {\bibfnamefont
  {R.}~\bibnamefont {Bruce}}, \bibinfo {author} {\bibfnamefont
  {D.}~\bibnamefont {Budker}}, \bibinfo {author} {\bibfnamefont
  {K.}~\bibnamefont {Cassou}}, \bibinfo {author} {\bibfnamefont
  {F.}~\bibnamefont {Castelli}}, \bibinfo {author} {\bibfnamefont
  {I.}~\bibnamefont {Chaikovska}}, \bibinfo {author} {\bibfnamefont
  {C.}~\bibnamefont {Curatolo}}, \bibinfo {author} {\bibfnamefont
  {P.}~\bibnamefont {Czodrowski}}, \bibinfo {author} {\bibfnamefont
  {D.}~\bibnamefont {Derevianko}}, \bibinfo {author} {\bibfnamefont
  {D.}~\bibnamefont {Dupraz}}, \bibinfo {author} {\bibfnamefont
  {Y.}~\bibnamefont {Dutheil}}, \bibinfo {author} {\bibfnamefont
  {K.}~\bibnamefont {Dzierzega}}, \bibinfo {author} {\bibfnamefont
  {V.}~\bibnamefont {Fedosseev}}, \bibinfo {author} {\bibfnamefont
  {N.}~\bibnamefont {Fuster~Martinez}}, \bibinfo {author} {\bibfnamefont
  {S.~M.}\ \bibnamefont {Gibson}}, \bibinfo {author} {\bibfnamefont
  {B.}~\bibnamefont {Goddard}}, \bibinfo {author} {\bibfnamefont
  {A.}~\bibnamefont {Gorzawski}}, \bibinfo {author} {\bibfnamefont
  {S.}~\bibnamefont {Hirlander}}, \bibinfo {author} {\bibfnamefont {J.~M.}\
  \bibnamefont {Jowett}}, \bibinfo {author} {\bibfnamefont {R.}~\bibnamefont
  {Kersevan}}, \bibinfo {author} {\bibfnamefont {M.}~\bibnamefont {Kowalska}},
  \bibinfo {author} {\bibfnamefont {F.}~\bibnamefont {Kroeger}}, \bibinfo
  {author} {\bibfnamefont {D.}~\bibnamefont {Kuchler}}, \bibinfo {author}
  {\bibfnamefont {M.}~\bibnamefont {Lamont}}, \bibinfo {author} {\bibfnamefont
  {T.}~\bibnamefont {Lefevre}}, \bibinfo {author} {\bibfnamefont
  {D.}~\bibnamefont {Manglunki}}, \bibinfo {author} {\bibfnamefont
  {B.}~\bibnamefont {Marsh}}, \bibinfo {author} {\bibfnamefont
  {A.}~\bibnamefont {Martens}}, \bibinfo {author} {\bibfnamefont
  {J.}~\bibnamefont {Molson}}, \bibinfo {author} {\bibfnamefont
  {D.}~\bibnamefont {Nutarelli}}, \bibinfo {author} {\bibfnamefont {L.~J.}\
  \bibnamefont {Nevay}}, \bibinfo {author} {\bibfnamefont {A.}~\bibnamefont
  {Petrenko}}, \bibinfo {author} {\bibfnamefont {V.}~\bibnamefont {Petrillo}},
  \bibinfo {author} {\bibfnamefont {W.}~\bibnamefont {Placzek}}, \bibinfo
  {author} {\bibfnamefont {S.}~\bibnamefont {Redaelli}}, \bibinfo {author}
  {\bibfnamefont {Y.}~\bibnamefont {Peinaud}}, \bibinfo {author} {\bibfnamefont
  {P.}~\bibnamefont {Pustelny}}, \bibinfo {author} {\bibfnamefont
  {S.}~\bibnamefont {Rochester}}, \bibinfo {author} {\bibfnamefont
  {M.}~\bibnamefont {Sapinski}}, \bibinfo {author} {\bibfnamefont
  {M.}~\bibnamefont {Schaumann}}, \bibinfo {author} {\bibfnamefont
  {R.}~\bibnamefont {Scrivens}}, \bibinfo {author} {\bibfnamefont
  {L.}~\bibnamefont {Serafini}}, \bibinfo {author} {\bibfnamefont {V.~P.}\
  \bibnamefont {Shevelko}}, \bibinfo {author} {\bibfnamefont {T.}~\bibnamefont
  {Stoehlker}}, \bibinfo {author} {\bibfnamefont {A.}~\bibnamefont
  {Surzhykov}}, \bibinfo {author} {\bibfnamefont {I.}~\bibnamefont
  {Tolstikhina}}, \bibinfo {author} {\bibfnamefont {F.}~\bibnamefont
  {Velotti}}, \bibinfo {author} {\bibfnamefont {G.}~\bibnamefont {Weber}},
  \bibinfo {author} {\bibfnamefont {Y.~K.}\ \bibnamefont {Wu}}, \bibinfo
  {author} {\bibfnamefont {C.}~\bibnamefont {Yin-Vallgren}}, \bibinfo {author}
  {\bibfnamefont {M.}~\bibnamefont {Zanetti}}, \bibinfo {author} {\bibfnamefont
  {F.}~\bibnamefont {Zimmermann}}, \bibinfo {author} {\bibfnamefont {M.~S.}\
  \bibnamefont {Zolotorev}},\ and\ \bibinfo {author} {\bibfnamefont
  {F.}~\bibnamefont {Zomer}} (\bibinfo {collaboration} {Gamma Factory Study
  Group Collaboration}),\ }\href {https://cds.cern.ch/record/2690736} {\emph
  {\bibinfo {title} {{Gamma Factory Proof-of-Principle Experiment}}}},\
  \bibinfo {type} {Tech. Rep.}\ \bibinfo {number} {CERN-SPSC-2019-031.
  SPSC-I-253}\ (\bibinfo  {institution} {CERN},\ \bibinfo {address} {Geneva},\
  \bibinfo {year} {2019})\BibitemShut {NoStop}%
\bibitem [{\citenamefont {Benedikt}\ \emph {et~al.}(2004)\citenamefont
  {Benedikt}, \citenamefont {Collier}, \citenamefont {Mertens}, \citenamefont
  {Poole},\ and\ \citenamefont {{K.~Schindl (editors)}}}]{lhcdesignV3}%
  \BibitemOpen
  \bibfield  {author} {\bibinfo {author} {\bibfnamefont {M.}~\bibnamefont
  {Benedikt}}, \bibinfo {author} {\bibfnamefont {P.}~\bibnamefont {Collier}},
  \bibinfo {author} {\bibfnamefont {V.}~\bibnamefont {Mertens}}, \bibinfo
  {author} {\bibfnamefont {J.}~\bibnamefont {Poole}},\ and\ \bibinfo {author}
  {\bibnamefont {{K.~Schindl (editors)}}},\ }\bibfield  {title} {\bibinfo
  {title} {{LHC} design report v.3 : The {LHC} injector chain},\ }\href@noop {}
  {\bibfield  {journal} {\bibinfo  {journal} {CERN-2004-003-V3}\ } (\bibinfo
  {year} {2004})}\BibitemShut {NoStop}%
\bibitem [{\citenamefont {Küchler}\ \emph {et~al.}(2014)\citenamefont
  {Küchler}, \citenamefont {O’Neil}, \citenamefont {Scrivens},\ and\
  \citenamefont {Thomae}}]{kuchler14}%
  \BibitemOpen
  \bibfield  {author} {\bibinfo {author} {\bibfnamefont {D.}~\bibnamefont
  {Küchler}}, \bibinfo {author} {\bibfnamefont {M.}~\bibnamefont {O’Neil}},
  \bibinfo {author} {\bibfnamefont {R.}~\bibnamefont {Scrivens}},\ and\
  \bibinfo {author} {\bibfnamefont {R.}~\bibnamefont {Thomae}},\ }\bibfield
  {title} {\bibinfo {title} {Preparation of a primary argon beam for the cern
  fixed target physics},\ }\bibfield  {journal} {\bibinfo  {journal} {Review of
  Scientific Instruments}\ }\textbf {\bibinfo {volume} {85}},\ \href
  {https://doi.org/10.1063/1.4854275} {10.1063/1.4854275} (\bibinfo {year}
  {2014})\BibitemShut {NoStop}%
\bibitem [{\citenamefont {Arduini}\ \emph {et~al.}(1996)\citenamefont
  {Arduini}, \citenamefont {Bailey}, \citenamefont {Bohl}, \citenamefont
  {Burkhardt}, \citenamefont {Cappi}, \citenamefont {Carter}, \citenamefont
  {Cornelis}, \citenamefont {Dach}, \citenamefont {de~Rijk}, \citenamefont
  {Faugier}, \citenamefont {Ferioli}, \citenamefont {Jakob}, \citenamefont
  {Jonker}, \citenamefont {Manglunki}, \citenamefont {Martini}, \citenamefont
  {Martini}, \citenamefont {Riunaud}, \citenamefont {Scheidenberger},
  \citenamefont {Vandorpe}, \citenamefont {Vos},\ and\ \citenamefont
  {Zanolli}}]{Arduini:308372}%
  \BibitemOpen
  \bibfield  {author} {\bibinfo {author} {\bibfnamefont {G.}~\bibnamefont
  {Arduini}}, \bibinfo {author} {\bibfnamefont {R.}~\bibnamefont {Bailey}},
  \bibinfo {author} {\bibfnamefont {T.}~\bibnamefont {Bohl}}, \bibinfo {author}
  {\bibfnamefont {H.}~\bibnamefont {Burkhardt}}, \bibinfo {author}
  {\bibfnamefont {R.}~\bibnamefont {Cappi}}, \bibinfo {author} {\bibfnamefont
  {C.}~\bibnamefont {Carter}}, \bibinfo {author} {\bibfnamefont
  {K.}~\bibnamefont {Cornelis}}, \bibinfo {author} {\bibfnamefont
  {M.}~\bibnamefont {Dach}}, \bibinfo {author} {\bibfnamefont {G.}~\bibnamefont
  {de~Rijk}}, \bibinfo {author} {\bibfnamefont {A.}~\bibnamefont {Faugier}},
  \bibinfo {author} {\bibfnamefont {G.}~\bibnamefont {Ferioli}}, \bibinfo
  {author} {\bibfnamefont {H.}~\bibnamefont {Jakob}}, \bibinfo {author}
  {\bibfnamefont {M.}~\bibnamefont {Jonker}}, \bibinfo {author} {\bibfnamefont
  {D.}~\bibnamefont {Manglunki}}, \bibinfo {author} {\bibfnamefont
  {G.}~\bibnamefont {Martini}}, \bibinfo {author} {\bibfnamefont
  {M.}~\bibnamefont {Martini}}, \bibinfo {author} {\bibfnamefont {J.~P.}\
  \bibnamefont {Riunaud}}, \bibinfo {author} {\bibfnamefont {C.}~\bibnamefont
  {Scheidenberger}}, \bibinfo {author} {\bibfnamefont {B.}~\bibnamefont
  {Vandorpe}}, \bibinfo {author} {\bibfnamefont {L.}~\bibnamefont {Vos}},\ and\
  \bibinfo {author} {\bibfnamefont {M.}~\bibnamefont {Zanolli}},\ }\bibfield
  {title} {\bibinfo {title} {{Lead ion beam emittance and transmission studies
  in the PS-SPS complex at CERN}},\ }\href {https://cds.cern.ch/record/308372}
  {\ ,\ \bibinfo {pages} {4 p} (\bibinfo {year} {1996})}\BibitemShut {NoStop}%
\bibitem [{\citenamefont {Schaumann}\ \emph
  {et~al.}(2019{\natexlab{a}})\citenamefont {Schaumann}, \citenamefont
  {Alemany-Fernandez}, \citenamefont {Bartosik}, \citenamefont {Bohl},
  \citenamefont {Bruce}, \citenamefont {Hemelsoet}, \citenamefont {Hirlaender},
  \citenamefont {Jowett}, \citenamefont {Kain}, \citenamefont {Krasny},
  \citenamefont {Molson}, \citenamefont {Papotti}, \citenamefont {Camillocci},
  \citenamefont {Timko},\ and\ \citenamefont
  {Wenninger}}]{Schaumann_2019_ipac_part_strip_ions}%
  \BibitemOpen
  \bibfield  {author} {\bibinfo {author} {\bibfnamefont {M.}~\bibnamefont
  {Schaumann}}, \bibinfo {author} {\bibfnamefont {R.}~\bibnamefont
  {Alemany-Fernandez}}, \bibinfo {author} {\bibfnamefont {H.}~\bibnamefont
  {Bartosik}}, \bibinfo {author} {\bibfnamefont {T.}~\bibnamefont {Bohl}},
  \bibinfo {author} {\bibfnamefont {R.}~\bibnamefont {Bruce}}, \bibinfo
  {author} {\bibfnamefont {G.~H.}\ \bibnamefont {Hemelsoet}}, \bibinfo {author}
  {\bibfnamefont {S.}~\bibnamefont {Hirlaender}}, \bibinfo {author}
  {\bibfnamefont {J.~M.}\ \bibnamefont {Jowett}}, \bibinfo {author}
  {\bibfnamefont {V.}~\bibnamefont {Kain}}, \bibinfo {author} {\bibfnamefont
  {M.~W.}\ \bibnamefont {Krasny}}, \bibinfo {author} {\bibfnamefont
  {J.}~\bibnamefont {Molson}}, \bibinfo {author} {\bibfnamefont
  {G.}~\bibnamefont {Papotti}}, \bibinfo {author} {\bibfnamefont {M.~S.}\
  \bibnamefont {Camillocci}}, \bibinfo {author} {\bibfnamefont
  {H.}~\bibnamefont {Timko}},\ and\ \bibinfo {author} {\bibfnamefont
  {J.}~\bibnamefont {Wenninger}},\ }\bibfield  {title} {\bibinfo {title} {First
  partially stripped ions in the {LHC} ($^{208}$pb$^{81+}$)},\ }\href
  {https://doi.org/10.1088/1742-6596/1350/1/012071} {\bibfield  {journal}
  {\bibinfo  {journal} {Journal of Physics: Conference Series}\ }\textbf
  {\bibinfo {volume} {1350}},\ \bibinfo {pages} {012071} (\bibinfo {year}
  {2019}{\natexlab{a}})}\BibitemShut {NoStop}%
\bibitem [{\citenamefont {Schaumann}\ \emph
  {et~al.}(2019{\natexlab{b}})\citenamefont {Schaumann}, \citenamefont
  {Abramov}, \citenamefont {Alemany~Fernandez}, \citenamefont {Argyropoulos},
  \citenamefont {Bartosik}, \citenamefont {Biancacci}, \citenamefont {Bohl},
  \citenamefont {Bracco}, \citenamefont {Bruce}, \citenamefont {Burger},
  \citenamefont {Cornelis}, \citenamefont {Fuster~Martinez}, \citenamefont
  {Goddard}, \citenamefont {Gorzawski}, \citenamefont {Giachino}, \citenamefont
  {Hemelsoet}, \citenamefont {Hirlander}, \citenamefont {Jebramcik},
  \citenamefont {Jowett}, \citenamefont {Kain}, \citenamefont {Krasny},
  \citenamefont {Molson}, \citenamefont {Papotti}, \citenamefont
  {Solfaroli~Camillocci}, \citenamefont {Timko}, \citenamefont {Valuch},
  \citenamefont {Velotti},\ and\ \citenamefont
  {Wenninger}}]{Schaumann:2670544}%
  \BibitemOpen
  \bibfield  {author} {\bibinfo {author} {\bibfnamefont {M.}~\bibnamefont
  {Schaumann}}, \bibinfo {author} {\bibfnamefont {A.}~\bibnamefont {Abramov}},
  \bibinfo {author} {\bibfnamefont {R.}~\bibnamefont {Alemany~Fernandez}},
  \bibinfo {author} {\bibfnamefont {T.}~\bibnamefont {Argyropoulos}}, \bibinfo
  {author} {\bibfnamefont {H.}~\bibnamefont {Bartosik}}, \bibinfo {author}
  {\bibfnamefont {N.}~\bibnamefont {Biancacci}}, \bibinfo {author}
  {\bibfnamefont {T.}~\bibnamefont {Bohl}}, \bibinfo {author} {\bibfnamefont
  {C.}~\bibnamefont {Bracco}}, \bibinfo {author} {\bibfnamefont
  {R.}~\bibnamefont {Bruce}}, \bibinfo {author} {\bibfnamefont
  {S.}~\bibnamefont {Burger}}, \bibinfo {author} {\bibfnamefont
  {K.}~\bibnamefont {Cornelis}}, \bibinfo {author} {\bibfnamefont
  {N.}~\bibnamefont {Fuster~Martinez}}, \bibinfo {author} {\bibfnamefont
  {B.}~\bibnamefont {Goddard}}, \bibinfo {author} {\bibfnamefont
  {A.}~\bibnamefont {Gorzawski}}, \bibinfo {author} {\bibfnamefont
  {R.}~\bibnamefont {Giachino}}, \bibinfo {author} {\bibfnamefont {G.-H.}\
  \bibnamefont {Hemelsoet}}, \bibinfo {author} {\bibfnamefont {S.}~\bibnamefont
  {Hirlander}}, \bibinfo {author} {\bibfnamefont {M.~A.}\ \bibnamefont
  {Jebramcik}}, \bibinfo {author} {\bibfnamefont {J.}~\bibnamefont {Jowett}},
  \bibinfo {author} {\bibfnamefont {V.}~\bibnamefont {Kain}}, \bibinfo {author}
  {\bibfnamefont {M.}~\bibnamefont {Krasny}}, \bibinfo {author} {\bibfnamefont
  {J.}~\bibnamefont {Molson}}, \bibinfo {author} {\bibfnamefont
  {G.}~\bibnamefont {Papotti}}, \bibinfo {author} {\bibfnamefont
  {M.}~\bibnamefont {Solfaroli~Camillocci}}, \bibinfo {author} {\bibfnamefont
  {H.}~\bibnamefont {Timko}}, \bibinfo {author} {\bibfnamefont
  {D.}~\bibnamefont {Valuch}}, \bibinfo {author} {\bibfnamefont {F.~M.}\
  \bibnamefont {Velotti}},\ and\ \bibinfo {author} {\bibfnamefont
  {J.}~\bibnamefont {Wenninger}},\ }\bibfield  {title} {\bibinfo {title}
  {{MD3284: Partially Stripped Ions in the LHC}},\ }\href
  {https://cds.cern.ch/record/2670544} {\  (\bibinfo {year}
  {2019}{\natexlab{b}})}\BibitemShut {NoStop}%
\bibitem [{\citenamefont {{R.W.~Assmann}}(2005)}]{assmann05chamonix}%
  \BibitemOpen
  \bibfield  {author} {\bibinfo {author} {\bibnamefont {{R.W.~Assmann}}},\
  }\bibfield  {title} {\bibinfo {title} {{Collimators and Beam Absorbers for
  Cleaning and Machine Protection}},\ }\href@noop {} {\bibfield  {journal}
  {\bibinfo  {journal} {Proceedings of the LHC Project Workshop - Chamonix XIV,
  Chamonix, France}\ ,\ \bibinfo {pages} {261}} (\bibinfo {year}
  {2005})}\BibitemShut {NoStop}%
\bibitem [{\citenamefont {{R.W.~Assmann \textit{et al.}}}(2006)}]{assmann06}%
  \BibitemOpen
  \bibfield  {author} {\bibinfo {author} {\bibnamefont {{R.W.~Assmann
  \textit{et al.}}}},\ }\bibfield  {title} {\bibinfo {title} {{The Final
  Collimation System for the {LHC}}},\ }\href@noop {} {\bibfield  {journal}
  {\bibinfo  {journal} {Proc. of the European Particle Accelerator Conference
  2006, Edinburgh, Scotland}\ ,\ \bibinfo {pages} {986}} (\bibinfo {year}
  {2006})}\BibitemShut {NoStop}%
\bibitem [{\citenamefont {Bruce}\ \emph
  {et~al.}(2014{\natexlab{a}})\citenamefont {Bruce} \emph
  {et~al.}}]{bruce14_PRSTAB_sixtr}%
  \BibitemOpen
  \bibfield  {author} {\bibinfo {author} {\bibfnamefont {R.}~\bibnamefont
  {Bruce}} \emph {et~al.},\ }\bibfield  {title} {\bibinfo {title} {{Simulations
  and measurements of beam loss patterns at the CERN Large Hadron Collider}},\
  }\href {https://doi.org/10.1103/PhysRevSTAB.17.081004} {\bibfield  {journal}
  {\bibinfo  {journal} {Phys. Rev. ST Accel. Beams}\ }\textbf {\bibinfo
  {volume} {17}},\ \bibinfo {pages} {081004} (\bibinfo {year}
  {2014}{\natexlab{a}})}\BibitemShut {NoStop}%
\bibitem [{\citenamefont {Bruce}\ \emph {et~al.}(2015)\citenamefont {Bruce},
  \citenamefont {Assmann},\ and\ \citenamefont
  {Redaelli}}]{bruce15_PRSTAB_betaStar}%
  \BibitemOpen
  \bibfield  {author} {\bibinfo {author} {\bibfnamefont {R.}~\bibnamefont
  {Bruce}}, \bibinfo {author} {\bibfnamefont {R.~W.}\ \bibnamefont {Assmann}},\
  and\ \bibinfo {author} {\bibfnamefont {S.}~\bibnamefont {Redaelli}},\
  }\bibfield  {title} {\bibinfo {title} {{Calculations of safe collimator
  settings and ${\ensuremath{\beta}}^{*}$ at the CERN Large Hadron Collider}},\
  }\href {https://doi.org/10.1103/PhysRevSTAB.18.061001} {\bibfield  {journal}
  {\bibinfo  {journal} {Phys. Rev. ST Accel. Beams}\ }\textbf {\bibinfo
  {volume} {18}},\ \bibinfo {pages} {061001} (\bibinfo {year}
  {2015})}\BibitemShut {NoStop}%
\bibitem [{\citenamefont {Valentino}\ \emph {et~al.}(2017)\citenamefont
  {Valentino}, \citenamefont {Baud}, \citenamefont {Bruce}, \citenamefont
  {Gasior}, \citenamefont {Mereghetti}, \citenamefont {Mirarchi}, \citenamefont
  {Olexa}, \citenamefont {Redaelli}, \citenamefont {Salvachua}, \citenamefont
  {Valloni},\ and\ \citenamefont {Wenninger}}]{valentino17_PRSTAB}%
  \BibitemOpen
  \bibfield  {author} {\bibinfo {author} {\bibfnamefont {G.}~\bibnamefont
  {Valentino}}, \bibinfo {author} {\bibfnamefont {G.}~\bibnamefont {Baud}},
  \bibinfo {author} {\bibfnamefont {R.}~\bibnamefont {Bruce}}, \bibinfo
  {author} {\bibfnamefont {M.}~\bibnamefont {Gasior}}, \bibinfo {author}
  {\bibfnamefont {A.}~\bibnamefont {Mereghetti}}, \bibinfo {author}
  {\bibfnamefont {D.}~\bibnamefont {Mirarchi}}, \bibinfo {author}
  {\bibfnamefont {J.}~\bibnamefont {Olexa}}, \bibinfo {author} {\bibfnamefont
  {S.}~\bibnamefont {Redaelli}}, \bibinfo {author} {\bibfnamefont
  {S.}~\bibnamefont {Salvachua}}, \bibinfo {author} {\bibfnamefont
  {A.}~\bibnamefont {Valloni}},\ and\ \bibinfo {author} {\bibfnamefont
  {J.}~\bibnamefont {Wenninger}},\ }\bibfield  {title} {\bibinfo {title} {Final
  implementation, commissioning, and performance of embedded collimator beam
  position monitors in the large hadron collider},\ }\href
  {https://doi.org/10.1103/PhysRevAccelBeams.20.081002} {\bibfield  {journal}
  {\bibinfo  {journal} {Phys. Rev. Accel. Beams}\ }\textbf {\bibinfo {volume}
  {20}},\ \bibinfo {pages} {081002} (\bibinfo {year} {2017})}\BibitemShut
  {NoStop}%
\bibitem [{\citenamefont {Redaelli}(2016)}]{redaelli_coll}%
  \BibitemOpen
  \bibfield  {author} {\bibinfo {author} {\bibfnamefont {S.}~\bibnamefont
  {Redaelli}},\ }\bibfield  {title} {\bibinfo {title} {Beam cleaning and
  collimation systems},\ }\href
  {https://e-publishing.cern.ch/index.php/CYR/article/view/243} {\bibfield
  {journal} {\bibinfo  {journal} {CERN Yellow Reports}\ }\textbf {\bibinfo
  {volume} {2}},\ \bibinfo {pages} {403} (\bibinfo {year} {2016})}\BibitemShut
  {NoStop}%
\bibitem [{\citenamefont {Bruce}\ \emph
  {et~al.}(2014{\natexlab{b}})\citenamefont {Bruce}, \citenamefont {Marsili},\
  and\ \citenamefont {Redaelli}}]{bruce14ipac_DS_coll}%
  \BibitemOpen
  \bibfield  {author} {\bibinfo {author} {\bibfnamefont {R.}~\bibnamefont
  {Bruce}}, \bibinfo {author} {\bibfnamefont {A.}~\bibnamefont {Marsili}},\
  and\ \bibinfo {author} {\bibfnamefont {S.}~\bibnamefont {Redaelli}},\
  }\bibfield  {title} {\bibinfo {title} {{Cleaning Performance with 11T Dipoles
  and Local Dispersion Suppressor Collimation at the LHC}},\ }\href
  {http://accelconf.web.cern.ch/AccelConf/IPAC2014/papers/mopro042.pdf}
  {\bibfield  {journal} {\bibinfo  {journal} {Proceedings of the International
  Particle Accelerator Conference 2014, Dresden, Germany}\ ,\ \bibinfo {pages}
  {170}} (\bibinfo {year} {2014}{\natexlab{b}})}\BibitemShut {NoStop}%
\bibitem [{\citenamefont {Jowett}\ \emph {et~al.}(2004)\citenamefont {Jowett},
  \citenamefont {Braun}, \citenamefont {Gresham}, \citenamefont {Mahner},
  \citenamefont {Nicholson},\ and\ \citenamefont {Shaposhnikova}}]{epac2004}%
  \BibitemOpen
  \bibfield  {author} {\bibinfo {author} {\bibfnamefont {J.~M.}\ \bibnamefont
  {Jowett}}, \bibinfo {author} {\bibfnamefont {H.~H.}\ \bibnamefont {Braun}},
  \bibinfo {author} {\bibfnamefont {M.~I.}\ \bibnamefont {Gresham}}, \bibinfo
  {author} {\bibfnamefont {E.}~\bibnamefont {Mahner}}, \bibinfo {author}
  {\bibfnamefont {A.~N.}\ \bibnamefont {Nicholson}},\ and\ \bibinfo {author}
  {\bibfnamefont {E.}~\bibnamefont {Shaposhnikova}},\ }\bibfield  {title}
  {\bibinfo {title} {{Limits to the Performance of the LHC with Ion Beams}},\
  }\href@noop {} {\bibfield  {journal} {\bibinfo  {journal} {Proc. of the
  European Particle Accelerator Conf. 2004, Lucerne}\ ,\ \bibinfo {pages}
  {578}} (\bibinfo {year} {2004})}\BibitemShut {NoStop}%
\bibitem [{\citenamefont {{P.D. Hermes \textit{et
  al.}}}(2016)}]{hermes16_ion_quench_test}%
  \BibitemOpen
  \bibfield  {author} {\bibinfo {author} {\bibnamefont {{P.D. Hermes \textit{et
  al.}}}},\ }\bibfield  {title} {\bibinfo {title} {{LHC Heavy-Ion Collimation
  Quench Test at 6.37Z TeV}},\ }\href {http://cds.cern.ch/record/2136828}
  {\bibfield  {journal} {\bibinfo  {journal} {CERN-ACC-NOTE-2016-0031}\ }
  (\bibinfo {year} {2016})}\BibitemShut {NoStop}%
\bibitem [{\citenamefont {Hermes}\ \emph
  {et~al.}(2016{\natexlab{a}})\citenamefont {Hermes}, \citenamefont {Bruce},
  \citenamefont {Jowett}, \citenamefont {Redaelli}, \citenamefont {Ferrando},
  \citenamefont {Valentino},\ and\ \citenamefont {Wollmann}}]{hermes16_nim}%
  \BibitemOpen
  \bibfield  {author} {\bibinfo {author} {\bibfnamefont {P.}~\bibnamefont
  {Hermes}}, \bibinfo {author} {\bibfnamefont {R.}~\bibnamefont {Bruce}},
  \bibinfo {author} {\bibfnamefont {J.}~\bibnamefont {Jowett}}, \bibinfo
  {author} {\bibfnamefont {S.}~\bibnamefont {Redaelli}}, \bibinfo {author}
  {\bibfnamefont {B.~S.}\ \bibnamefont {Ferrando}}, \bibinfo {author}
  {\bibfnamefont {G.}~\bibnamefont {Valentino}},\ and\ \bibinfo {author}
  {\bibfnamefont {D.}~\bibnamefont {Wollmann}},\ }\bibfield  {title} {\bibinfo
  {title} {Measured and simulated heavy-ion beam loss patterns at the {CERN
  Large Hadron Collider}},\ }\href
  {https://doi.org/http://dx.doi.org/10.1016/j.nima.2016.02.050} {\bibfield
  {journal} {\bibinfo  {journal} {Nucl. Instrum. Methods Phys. Res. A}\
  }\textbf {\bibinfo {volume} {819}},\ \bibinfo {pages} {73 } (\bibinfo {year}
  {2016}{\natexlab{a}})}\BibitemShut {NoStop}%
\bibitem [{\citenamefont {Dehning}(2016)}]{blmSystem}%
  \BibitemOpen
  \bibfield  {author} {\bibinfo {author} {\bibfnamefont {B.}~\bibnamefont
  {Dehning}},\ }\bibfield  {title} {\bibinfo {title} {Beam loss monitors at
  lhc},\ }\href {https://e-publishing.cern.ch/index.php/CYR/article/view/238}
  {\bibfield  {journal} {\bibinfo  {journal} {CERN Yellow Reports}\ }\textbf
  {\bibinfo {volume} {2}},\ \bibinfo {pages} {303} (\bibinfo {year}
  {2016})}\BibitemShut {NoStop}%
\bibitem [{\citenamefont {Holzer}\ \emph {et~al.}(2005)\citenamefont {Holzer},
  \citenamefont {Dehning}, \citenamefont {Effinger}, \citenamefont {Emery},
  \citenamefont {Ferioli}, \citenamefont {Gonzalez}, \citenamefont
  {Gschwendtner}, \citenamefont {Guaglio}, \citenamefont {Hodgson},
  \citenamefont {Kramer}, \citenamefont {Leitner}, \citenamefont {Ponce},
  \citenamefont {Prieto}, \citenamefont {Stockner},\ and\ \citenamefont
  {Zamantzas}}]{holzer05}%
  \BibitemOpen
  \bibfield  {author} {\bibinfo {author} {\bibfnamefont {E.}~\bibnamefont
  {Holzer}}, \bibinfo {author} {\bibfnamefont {B.}~\bibnamefont {Dehning}},
  \bibinfo {author} {\bibfnamefont {E.}~\bibnamefont {Effinger}}, \bibinfo
  {author} {\bibfnamefont {J.}~\bibnamefont {Emery}}, \bibinfo {author}
  {\bibfnamefont {G.}~\bibnamefont {Ferioli}}, \bibinfo {author} {\bibfnamefont
  {J.}~\bibnamefont {Gonzalez}}, \bibinfo {author} {\bibfnamefont
  {E.}~\bibnamefont {Gschwendtner}}, \bibinfo {author} {\bibfnamefont
  {G.}~\bibnamefont {Guaglio}}, \bibinfo {author} {\bibfnamefont
  {M.}~\bibnamefont {Hodgson}}, \bibinfo {author} {\bibfnamefont
  {D.}~\bibnamefont {Kramer}}, \bibinfo {author} {\bibfnamefont
  {R.}~\bibnamefont {Leitner}}, \bibinfo {author} {\bibfnamefont
  {L.}~\bibnamefont {Ponce}}, \bibinfo {author} {\bibfnamefont
  {V.}~\bibnamefont {Prieto}}, \bibinfo {author} {\bibfnamefont
  {M.}~\bibnamefont {Stockner}},\ and\ \bibinfo {author} {\bibfnamefont
  {C.}~\bibnamefont {Zamantzas}},\ }\bibfield  {title} {\bibinfo {title} {{Beam
  Loss Monitoring System for the LHC}},\ }\href
  {https://doi.org/10.1109/NSSMIC.2005.1596433} {\bibfield  {journal} {\bibinfo
   {journal} {IEEE Nuclear Science Symposium Conference Record}\ }\textbf
  {\bibinfo {volume} {2}},\ \bibinfo {pages} {1052} (\bibinfo {year}
  {2005})}\BibitemShut {NoStop}%
\bibitem [{\citenamefont {{E. B.~Holzer \textit{et al}}}(2008)}]{holzer08a}%
  \BibitemOpen
  \bibfield  {author} {\bibinfo {author} {\bibnamefont {{E. B.~Holzer
  \textit{et al}}}},\ }\bibfield  {title} {\bibinfo {title} {Development,
  production and testing of 4500 beam loss monitors},\ }\href@noop {}
  {\bibfield  {journal} {\bibinfo  {journal} {Proc. of the European Particle
  Accelerator Conf. 2008, Genoa, Italy}\ ,\ \bibinfo {pages} {1134}} (\bibinfo
  {year} {2008})}\BibitemShut {NoStop}%
\bibitem [{\citenamefont {Bruce}\ \emph {et~al.}(2020)\citenamefont {Bruce},
  \citenamefont {Argyropoulos}, \citenamefont {Bartosik}, \citenamefont
  {Maria}, \citenamefont {Fuster-Martinez}, \citenamefont {Jebramcik},
  \citenamefont {Jowett}, \citenamefont {Mounet}, \citenamefont {Redaelli},
  \citenamefont {Rumolo}, \citenamefont {Schaumann},\ and\ \citenamefont
  {Timko}}]{bruce20_HL_ion_report}%
  \BibitemOpen
  \bibfield  {author} {\bibinfo {author} {\bibfnamefont {R.}~\bibnamefont
  {Bruce}}, \bibinfo {author} {\bibfnamefont {T.}~\bibnamefont {Argyropoulos}},
  \bibinfo {author} {\bibfnamefont {H.}~\bibnamefont {Bartosik}}, \bibinfo
  {author} {\bibfnamefont {R.~D.}\ \bibnamefont {Maria}}, \bibinfo {author}
  {\bibfnamefont {N.}~\bibnamefont {Fuster-Martinez}}, \bibinfo {author}
  {\bibfnamefont {M.}~\bibnamefont {Jebramcik}}, \bibinfo {author}
  {\bibfnamefont {J.}~\bibnamefont {Jowett}}, \bibinfo {author} {\bibfnamefont
  {N.}~\bibnamefont {Mounet}}, \bibinfo {author} {\bibfnamefont
  {S.}~\bibnamefont {Redaelli}}, \bibinfo {author} {\bibfnamefont
  {G.}~\bibnamefont {Rumolo}}, \bibinfo {author} {\bibfnamefont
  {M.}~\bibnamefont {Schaumann}},\ and\ \bibinfo {author} {\bibfnamefont
  {H.}~\bibnamefont {Timko}},\ }\bibfield  {title} {\bibinfo {title} {Hl-lhc
  operational scenario for pb-pb and p-pb operation},\ }\href@noop {}
  {\bibfield  {journal} {\bibinfo  {journal} {CERN report in preparation}\ }
  (\bibinfo {year} {2020})}\BibitemShut {NoStop}%
\bibitem [{\citenamefont {Tolstikhina}\ and\ \citenamefont
  {Shevelko}(2018)}]{Tolstikhina_2018}%
  \BibitemOpen
  \bibfield  {author} {\bibinfo {author} {\bibfnamefont {I.~Y.}\ \bibnamefont
  {Tolstikhina}}\ and\ \bibinfo {author} {\bibfnamefont {V.~P.}\ \bibnamefont
  {Shevelko}},\ }\bibfield  {title} {\bibinfo {title} {Influence of atomic
  processes on charge states and fractions of fast heavy ions passing through
  gaseous, solid, and plasma targets},\ }\href
  {https://doi.org/10.3367/ufne.2017.02.038071} {\bibfield  {journal} {\bibinfo
   {journal} {Physics-Uspekhi}\ }\textbf {\bibinfo {volume} {61}},\ \bibinfo
  {pages} {247} (\bibinfo {year} {2018})}\BibitemShut {NoStop}%
\bibitem [{\citenamefont {Herr}\ and\ \citenamefont
  {Schmidt}(2004{\natexlab{a}})}]{herr04}%
  \BibitemOpen
  \bibfield  {author} {\bibinfo {author} {\bibfnamefont {W.}~\bibnamefont
  {Herr}}\ and\ \bibinfo {author} {\bibfnamefont {F.}~\bibnamefont {Schmidt}},\
  }\bibfield  {title} {\bibinfo {title} {A {MAD-X} primer},\ }\href@noop {}
  {\bibfield  {journal} {\bibinfo  {journal} {CERN-AB-2004-027-AB}\ } (\bibinfo
  {year} {2004}{\natexlab{a}})}\BibitemShut {NoStop}%
\bibitem [{\citenamefont {Herr}\ and\ \citenamefont
  {Schmidt}(2004{\natexlab{b}})}]{madx}%
  \BibitemOpen
  \bibfield  {author} {\bibinfo {author} {\bibfnamefont {W.}~\bibnamefont
  {Herr}}\ and\ \bibinfo {author} {\bibfnamefont {F.}~\bibnamefont {Schmidt}},\
  }\bibfield  {title} {\bibinfo {title} {{A MAD-X Primer}},\ }\href@noop {} {\
  ,\ \bibinfo {pages} {32 p} (\bibinfo {year}
  {2004}{\natexlab{b}})}\BibitemShut {NoStop}%
\bibitem [{\citenamefont {Schmidt}(1994)}]{schmidt94}%
  \BibitemOpen
  \bibfield  {author} {\bibinfo {author} {\bibfnamefont {F.}~\bibnamefont
  {Schmidt}},\ }\bibfield  {title} {\bibinfo {title} {{SixTrack. User's
  Reference Manual}},\ }\href@noop {} {\bibfield  {journal} {\bibinfo
  {journal} {CERN/SL/94-56-AP}\ } (\bibinfo {year} {1994})}\BibitemShut
  {NoStop}%
\bibitem [{six()}]{sixtrack-web}%
  \BibitemOpen
  \href@noop {} {\bibinfo {title} {Sixtrack web site}},\ \bibinfo
  {howpublished} {\url{http://sixtrack.web.cern.ch/SixTrack/}},\ \bibinfo
  {note} {{\tt SixTrack} Web site, {\sf
  http://sixtrack.web.cern.ch/SixTrack/}}\BibitemShut {NoStop}%
\bibitem [{\citenamefont {Bohlen}\ \emph {et~al.}(2014)\citenamefont {Bohlen},
  \citenamefont {Cerutti}, \citenamefont {Chin}, \citenamefont {Fass\`{o}},
  \citenamefont {Ferrari}, \citenamefont {Ortega}, \citenamefont {Mairani},
  \citenamefont {Sala}, \citenamefont {Smirnov},\ and\ \citenamefont
  {Vlachoudis}}]{fluka14}%
  \BibitemOpen
  \bibfield  {author} {\bibinfo {author} {\bibfnamefont {T.}~\bibnamefont
  {Bohlen}}, \bibinfo {author} {\bibfnamefont {F.}~\bibnamefont {Cerutti}},
  \bibinfo {author} {\bibfnamefont {M.}~\bibnamefont {Chin}}, \bibinfo {author}
  {\bibfnamefont {A.}~\bibnamefont {Fass\`{o}}}, \bibinfo {author}
  {\bibfnamefont {A.}~\bibnamefont {Ferrari}}, \bibinfo {author} {\bibfnamefont
  {P.}~\bibnamefont {Ortega}}, \bibinfo {author} {\bibfnamefont
  {A.}~\bibnamefont {Mairani}}, \bibinfo {author} {\bibfnamefont
  {P.}~\bibnamefont {Sala}}, \bibinfo {author} {\bibfnamefont {G.}~\bibnamefont
  {Smirnov}},\ and\ \bibinfo {author} {\bibfnamefont {V.}~\bibnamefont
  {Vlachoudis}},\ }\bibfield  {title} {\bibinfo {title} {The {FLUKA} code:
  Developments and challenges for high energy and medical applications},\
  }\href@noop {} {\bibfield  {journal} {\bibinfo  {journal} {Nuclear Data
  Sheets}\ ,\ \bibinfo {pages} {211}} (\bibinfo {year} {2014})}\BibitemShut
  {NoStop}%
\bibitem [{\citenamefont {Battistoni}\ \emph {et~al.}(2015)\citenamefont
  {Battistoni} \emph {et~al.}}]{Battistoni:2015epi}%
  \BibitemOpen
  \bibfield  {author} {\bibinfo {author} {\bibfnamefont {G.}~\bibnamefont
  {Battistoni}} \emph {et~al.},\ }\bibfield  {title} {\bibinfo {title}
  {Overview of the fluka code},\ }\bibfield  {booktitle} {\emph {\bibinfo
  {booktitle} {{Proceedings, Joint International Conference on Supercomputing
  in Nuclear Applications + Monte Carlo (SNA + MC 2013): Paris, France, October
  27-31, 2013}}},\ }\href {https://doi.org/10.1016/j.anucene.2014.11.007}
  {\bibfield  {journal} {\bibinfo  {journal} {Annals Nucl. Energy}\ }\textbf
  {\bibinfo {volume} {82}},\ \bibinfo {pages} {10} (\bibinfo {year}
  {2015})}\BibitemShut {NoStop}%
\bibitem [{\citenamefont {{A. Mereghetti \textit{et
  al.}}}(2013)}]{mereghetti13_ipac}%
  \BibitemOpen
  \bibfield  {author} {\bibinfo {author} {\bibnamefont {{A. Mereghetti
  \textit{et al.}}}},\ }\bibfield  {title} {\bibinfo {title} {{Sixtrack-FLUKA
  active coupling for the upgrade of the SPS scrapers}},\ }\href@noop {}
  {\bibfield  {journal} {\bibinfo  {journal} {Proceedings of the International
  Particle Accelerator Conference 2013, Shanghai, China}\ ,\ \bibinfo {pages}
  {2657}} (\bibinfo {year} {2013})}\BibitemShut {NoStop}%
\bibitem [{\citenamefont {Skordis}\ \emph {et~al.}(2018)\citenamefont
  {Skordis}, \citenamefont {Vlachoudis}, \citenamefont {Bruce}, \citenamefont
  {Cerutti}, \citenamefont {Ferrari}, \citenamefont {Lechner}, \citenamefont
  {Mereghetti}, \citenamefont {Ortega}, \citenamefont {Redaelli},\ and\
  \citenamefont {Pastor}}]{skordis18_tracking_workshop}%
  \BibitemOpen
  \bibfield  {author} {\bibinfo {author} {\bibfnamefont {E.}~\bibnamefont
  {Skordis}}, \bibinfo {author} {\bibfnamefont {V.}~\bibnamefont {Vlachoudis}},
  \bibinfo {author} {\bibfnamefont {R.}~\bibnamefont {Bruce}}, \bibinfo
  {author} {\bibfnamefont {F.}~\bibnamefont {Cerutti}}, \bibinfo {author}
  {\bibfnamefont {A.}~\bibnamefont {Ferrari}}, \bibinfo {author} {\bibfnamefont
  {A.}~\bibnamefont {Lechner}}, \bibinfo {author} {\bibfnamefont
  {A.}~\bibnamefont {Mereghetti}}, \bibinfo {author} {\bibfnamefont
  {P.}~\bibnamefont {Ortega}}, \bibinfo {author} {\bibfnamefont
  {S.}~\bibnamefont {Redaelli}},\ and\ \bibinfo {author} {\bibfnamefont
  {D.~S.}\ \bibnamefont {Pastor}},\ }\bibfield  {title} {\bibinfo {title}
  {{FLUKA coupling to Sixtrack}},\ }\href
  {https://cds.cern.ch/record/2646800?ln=en} {\bibfield  {journal} {\bibinfo
  {journal} {CERN-2018-011-CP, Proceedings of the ICFA Mini-Workshop on
  Tracking for Collimation, CERN, Geneva, Switzerland}\ ,\ \bibinfo {pages}
  {17}} (\bibinfo {year} {2018})}\BibitemShut {NoStop}%
\bibitem [{\citenamefont {Hermes}(2016)}]{pascal-thesis}%
  \BibitemOpen
  \bibfield  {author} {\bibinfo {author} {\bibfnamefont {P.}~\bibnamefont
  {Hermes}},\ }\emph {\bibinfo {title} {{Heavy-Ion Collimation at the Large
  Hadron Collider : Simulations and Measurements}}},\ \href@noop {} {Ph.D.
  thesis},\ \bibinfo  {school} {University of Munster} (\bibinfo {year}
  {2016})\BibitemShut {NoStop}%
\bibitem [{\citenamefont {Auchmann}\ \emph {et~al.}(2015)\citenamefont
  {Auchmann}, \citenamefont {Baer}, \citenamefont {Bednarek}, \citenamefont
  {Bellodi}, \citenamefont {Bracco}, \citenamefont {Bruce}, \citenamefont
  {Cerutti}, \citenamefont {Chetvertkova}, \citenamefont {Dehning},
  \citenamefont {Granieri}, \citenamefont {Hofle}, \citenamefont {Holzer},
  \citenamefont {Lechner}, \citenamefont {Nebot Del~Busto}, \citenamefont
  {Priebe}, \citenamefont {Redaelli}, \citenamefont {Salvachua}, \citenamefont
  {Sapinski}, \citenamefont {Schmidt}, \citenamefont {Shetty}, \citenamefont
  {Skordis}, \citenamefont {Solfaroli}, \citenamefont {Steckert}, \citenamefont
  {Valuch}, \citenamefont {Verweij}, \citenamefont {Wenninger}, \citenamefont
  {Wollmann},\ and\ \citenamefont {Zerlauth}}]{auchmann15_PRSTAB}%
  \BibitemOpen
  \bibfield  {author} {\bibinfo {author} {\bibfnamefont {B.}~\bibnamefont
  {Auchmann}}, \bibinfo {author} {\bibfnamefont {T.}~\bibnamefont {Baer}},
  \bibinfo {author} {\bibfnamefont {M.}~\bibnamefont {Bednarek}}, \bibinfo
  {author} {\bibfnamefont {G.}~\bibnamefont {Bellodi}}, \bibinfo {author}
  {\bibfnamefont {C.}~\bibnamefont {Bracco}}, \bibinfo {author} {\bibfnamefont
  {R.}~\bibnamefont {Bruce}}, \bibinfo {author} {\bibfnamefont
  {F.}~\bibnamefont {Cerutti}}, \bibinfo {author} {\bibfnamefont
  {V.}~\bibnamefont {Chetvertkova}}, \bibinfo {author} {\bibfnamefont
  {B.}~\bibnamefont {Dehning}}, \bibinfo {author} {\bibfnamefont {P.~P.}\
  \bibnamefont {Granieri}}, \bibinfo {author} {\bibfnamefont {W.}~\bibnamefont
  {Hofle}}, \bibinfo {author} {\bibfnamefont {E.~B.}\ \bibnamefont {Holzer}},
  \bibinfo {author} {\bibfnamefont {A.}~\bibnamefont {Lechner}}, \bibinfo
  {author} {\bibfnamefont {E.}~\bibnamefont {Nebot Del~Busto}}, \bibinfo
  {author} {\bibfnamefont {A.}~\bibnamefont {Priebe}}, \bibinfo {author}
  {\bibfnamefont {S.}~\bibnamefont {Redaelli}}, \bibinfo {author}
  {\bibfnamefont {B.}~\bibnamefont {Salvachua}}, \bibinfo {author}
  {\bibfnamefont {M.}~\bibnamefont {Sapinski}}, \bibinfo {author}
  {\bibfnamefont {R.}~\bibnamefont {Schmidt}}, \bibinfo {author} {\bibfnamefont
  {N.}~\bibnamefont {Shetty}}, \bibinfo {author} {\bibfnamefont
  {E.}~\bibnamefont {Skordis}}, \bibinfo {author} {\bibfnamefont
  {M.}~\bibnamefont {Solfaroli}}, \bibinfo {author} {\bibfnamefont
  {J.}~\bibnamefont {Steckert}}, \bibinfo {author} {\bibfnamefont
  {D.}~\bibnamefont {Valuch}}, \bibinfo {author} {\bibfnamefont
  {A.}~\bibnamefont {Verweij}}, \bibinfo {author} {\bibfnamefont
  {J.}~\bibnamefont {Wenninger}}, \bibinfo {author} {\bibfnamefont
  {D.}~\bibnamefont {Wollmann}},\ and\ \bibinfo {author} {\bibfnamefont
  {M.}~\bibnamefont {Zerlauth}},\ }\bibfield  {title} {\bibinfo {title}
  {Testing beam-induced quench levels of lhc superconducting magnets},\ }\href
  {https://doi.org/10.1103/PhysRevSTAB.18.061002} {\bibfield  {journal}
  {\bibinfo  {journal} {Phys. Rev. ST Accel. Beams}\ }\textbf {\bibinfo
  {volume} {18}},\ \bibinfo {pages} {061002} (\bibinfo {year}
  {2015})}\BibitemShut {NoStop}%
\bibitem [{\citenamefont {Bruce}\ \emph {et~al.}(2017)\citenamefont {Bruce},
  \citenamefont {Bracco}, \citenamefont {Maria}, \citenamefont {Giovannozzi},
  \citenamefont {Mereghetti}, \citenamefont {Mirarchi}, \citenamefont
  {Redaelli}, \citenamefont {Quaranta},\ and\ \citenamefont
  {Salvachua}}]{bruce17_NIM_beta40cm}%
  \BibitemOpen
  \bibfield  {author} {\bibinfo {author} {\bibfnamefont {R.}~\bibnamefont
  {Bruce}}, \bibinfo {author} {\bibfnamefont {C.}~\bibnamefont {Bracco}},
  \bibinfo {author} {\bibfnamefont {R.~D.}\ \bibnamefont {Maria}}, \bibinfo
  {author} {\bibfnamefont {M.}~\bibnamefont {Giovannozzi}}, \bibinfo {author}
  {\bibfnamefont {A.}~\bibnamefont {Mereghetti}}, \bibinfo {author}
  {\bibfnamefont {D.}~\bibnamefont {Mirarchi}}, \bibinfo {author}
  {\bibfnamefont {S.}~\bibnamefont {Redaelli}}, \bibinfo {author}
  {\bibfnamefont {E.}~\bibnamefont {Quaranta}},\ and\ \bibinfo {author}
  {\bibfnamefont {B.}~\bibnamefont {Salvachua}},\ }\bibfield  {title} {\bibinfo
  {title} {Reaching record-low $\beta^*$ at the {CERN Large Hadron Collider}
  using a novel scheme of collimator settings and optics},\ }\href
  {https://doi.org/http://dx.doi.org/10.1016/j.nima.2016.12.039} {\bibfield
  {journal} {\bibinfo  {journal} {Nucl. Instrum. Methods Phys. Res. A}\
  }\textbf {\bibinfo {volume} {848}},\ \bibinfo {pages} {19 } (\bibinfo {year}
  {2017})}\BibitemShut {NoStop}%
\bibitem [{\citenamefont {Bruce}\ \emph {et~al.}(2019)\citenamefont {Bruce},
  \citenamefont {Huhtinen}, \citenamefont {Manousos}, \citenamefont {Cerutti},
  \citenamefont {Esposito}, \citenamefont {Kwee-Hinzmann}, \citenamefont
  {Lechner}, \citenamefont {Mereghetti}, \citenamefont {Mirarchi},
  \citenamefont {Redaelli},\ and\ \citenamefont
  {Salvachua}}]{bruce19_PRAB_beam-halo_backgrounds_ATLAS}%
  \BibitemOpen
  \bibfield  {author} {\bibinfo {author} {\bibfnamefont {R.}~\bibnamefont
  {Bruce}}, \bibinfo {author} {\bibfnamefont {M.}~\bibnamefont {Huhtinen}},
  \bibinfo {author} {\bibfnamefont {A.}~\bibnamefont {Manousos}}, \bibinfo
  {author} {\bibfnamefont {F.}~\bibnamefont {Cerutti}}, \bibinfo {author}
  {\bibfnamefont {L.}~\bibnamefont {Esposito}}, \bibinfo {author}
  {\bibfnamefont {R.}~\bibnamefont {Kwee-Hinzmann}}, \bibinfo {author}
  {\bibfnamefont {A.}~\bibnamefont {Lechner}}, \bibinfo {author} {\bibfnamefont
  {A.}~\bibnamefont {Mereghetti}}, \bibinfo {author} {\bibfnamefont
  {D.}~\bibnamefont {Mirarchi}}, \bibinfo {author} {\bibfnamefont
  {S.}~\bibnamefont {Redaelli}},\ and\ \bibinfo {author} {\bibfnamefont
  {B.}~\bibnamefont {Salvachua}},\ }\bibfield  {title} {\bibinfo {title}
  {Collimation-induced experimental background studies at the {CERN Large
  Hadron Collider}},\ }\href
  {https://doi.org/10.1103/PhysRevAccelBeams.22.021004} {\bibfield  {journal}
  {\bibinfo  {journal} {Phys. Rev. Accel. Beams}\ }\textbf {\bibinfo {volume}
  {22}},\ \bibinfo {pages} {021004} (\bibinfo {year} {2019})}\BibitemShut
  {NoStop}%
\bibitem [{\citenamefont {Hermes}\ \emph
  {et~al.}(2016{\natexlab{b}})\citenamefont {Hermes}, \citenamefont {Bruce},
  \citenamefont {Cerutti}, \citenamefont {Ferrari}, \citenamefont {Jowett},
  \citenamefont {Lechner}, \citenamefont {Mereghetti}, \citenamefont
  {Mirarchi}, \citenamefont {Ortega}, \citenamefont {Redaelli}, \citenamefont
  {Salvachua}, \citenamefont {Skordis}, \citenamefont {Valentino},\ and\
  \citenamefont {Vlachoudis}}]{hermes16_ipac_coupling}%
  \BibitemOpen
  \bibfield  {author} {\bibinfo {author} {\bibfnamefont {P.}~\bibnamefont
  {Hermes}}, \bibinfo {author} {\bibfnamefont {R.}~\bibnamefont {Bruce}},
  \bibinfo {author} {\bibfnamefont {F.}~\bibnamefont {Cerutti}}, \bibinfo
  {author} {\bibfnamefont {A.}~\bibnamefont {Ferrari}}, \bibinfo {author}
  {\bibfnamefont {J.}~\bibnamefont {Jowett}}, \bibinfo {author} {\bibfnamefont
  {A.}~\bibnamefont {Lechner}}, \bibinfo {author} {\bibfnamefont
  {A.}~\bibnamefont {Mereghetti}}, \bibinfo {author} {\bibfnamefont
  {D.}~\bibnamefont {Mirarchi}}, \bibinfo {author} {\bibfnamefont
  {P.}~\bibnamefont {Ortega}}, \bibinfo {author} {\bibfnamefont
  {S.}~\bibnamefont {Redaelli}}, \bibinfo {author} {\bibfnamefont
  {B.}~\bibnamefont {Salvachua}}, \bibinfo {author} {\bibfnamefont
  {E.}~\bibnamefont {Skordis}}, \bibinfo {author} {\bibfnamefont
  {G.}~\bibnamefont {Valentino}},\ and\ \bibinfo {author} {\bibfnamefont
  {V.}~\bibnamefont {Vlachoudis}},\ }\bibfield  {title} {\bibinfo {title}
  {{Simulation of Heavy-Ion Beam Losses with the SixTrack-FLUKA Active
  Coupling}},\ }\href
  {http://accelconf.web.cern.ch/AccelConf/ipac2016/papers/wepmw029.pdf}
  {\bibfield  {journal} {\bibinfo  {journal} {Proceedings of the International
  Particle Accelerator Conference 2016, Busan, Korea}\ } (\bibinfo {year}
  {2016}{\natexlab{b}})}\BibitemShut {NoStop}%
\bibitem [{\citenamefont {Apollinari}\ \emph {et~al.}(2017)\citenamefont
  {Apollinari}, \citenamefont {Alonso}, \citenamefont {Bruning}, \citenamefont
  {Fessia}, \citenamefont {Lamont}, \citenamefont {Rossi},\ and\ \citenamefont
  {{L. Tavian (editors)}}}]{hl-lhc-tech-design}%
  \BibitemOpen
  \bibfield  {author} {\bibinfo {author} {\bibfnamefont {G.}~\bibnamefont
  {Apollinari}}, \bibinfo {author} {\bibfnamefont {I.~B.}\ \bibnamefont
  {Alonso}}, \bibinfo {author} {\bibfnamefont {O.}~\bibnamefont {Bruning}},
  \bibinfo {author} {\bibfnamefont {P.}~\bibnamefont {Fessia}}, \bibinfo
  {author} {\bibfnamefont {M.}~\bibnamefont {Lamont}}, \bibinfo {author}
  {\bibfnamefont {L.}~\bibnamefont {Rossi}},\ and\ \bibinfo {author}
  {\bibnamefont {{L. Tavian (editors)}}},\ }\href
  {https://doi.org/http://dx.doi.org/10.23731/CYRM-2017-004} {\emph {\bibinfo
  {title} {{High-Luminosity Large Hadron Collider (HL-LHC): Technical Design
  Report V. 0.1}}}},\ CERN Yellow Reports: Monographs. CERN-2017-007-M\
  (\bibinfo  {publisher} {CERN},\ \bibinfo {address} {Geneva},\ \bibinfo {year}
  {2017})\BibitemShut {NoStop}%
\bibitem [{\citenamefont {Redaelli}\ \emph {et~al.}(2014)\citenamefont
  {Redaelli}, \citenamefont {Bertarelli}, \citenamefont {Bruce}, \citenamefont
  {Efthymiopoulos}, \citenamefont {Lechner}, ,\ and\ \citenamefont
  {Uythoven}}]{redaelli14_chamonix}%
  \BibitemOpen
  \bibfield  {author} {\bibinfo {author} {\bibfnamefont {S.}~\bibnamefont
  {Redaelli}}, \bibinfo {author} {\bibfnamefont {A.}~\bibnamefont
  {Bertarelli}}, \bibinfo {author} {\bibfnamefont {R.}~\bibnamefont {Bruce}},
  \bibinfo {author} {\bibfnamefont {I.}~\bibnamefont {Efthymiopoulos}},
  \bibinfo {author} {\bibfnamefont {A.}~\bibnamefont {Lechner}}, ,\ and\
  \bibinfo {author} {\bibfnamefont {J.}~\bibnamefont {Uythoven}},\ }\bibfield
  {title} {\bibinfo {title} {Collimation upgrades for {HL-LHC}},\ }\href
  {https://indico.cern.ch/event/315665/session/6/contribution/29/material/paper/0.pdf}
  {\bibfield  {journal} {\bibinfo  {journal} {Proceedings of the LHC
  Performance Workshop (Chamonix 2014), Chamonix, France}\ } (\bibinfo {year}
  {2014})}\BibitemShut {NoStop}%
\bibitem [{\citenamefont {Hermes}\ \emph {et~al.}(2015)\citenamefont {Hermes},
  \citenamefont {Bruce}, \citenamefont {Jowett},\ and\ \citenamefont
  {Redaelli}}]{hermes15_ipac}%
  \BibitemOpen
  \bibfield  {author} {\bibinfo {author} {\bibfnamefont {P.}~\bibnamefont
  {Hermes}}, \bibinfo {author} {\bibfnamefont {R.}~\bibnamefont {Bruce}},
  \bibinfo {author} {\bibfnamefont {J.}~\bibnamefont {Jowett}},\ and\ \bibinfo
  {author} {\bibfnamefont {S.}~\bibnamefont {Redaelli}},\ }\bibfield  {title}
  {\bibinfo {title} {{Betatron Cleaning for Heavy Ion Beams with IR7 Dispersion
  Suppressor Collimators}},\ }\href
  {http://accelconf.web.cern.ch/AccelConf/IPAC2015/papers/tupty025.pdf}
  {\bibfield  {journal} {\bibinfo  {journal} {Proceedings of the International
  Particle Accelerator Conference 2015, Richmond, VA, USA}\ } (\bibinfo {year}
  {2015})}\BibitemShut {NoStop}%
\bibitem [{\citenamefont {Zlobin}\ and\ \citenamefont
  {Schoerling}(2019)}]{Zlobin:2019ven}%
  \BibitemOpen
  \bibfield  {author} {\bibinfo {author} {\bibfnamefont {A.~V.}\ \bibnamefont
  {Zlobin}}\ and\ \bibinfo {author} {\bibfnamefont {D.}~\bibnamefont
  {Schoerling}},\ }\bibinfo {title} {{Superconducting Magnets for
  Accelerators}},\ in\ \href {https://doi.org/10.1007/978-3-030-16118-7\_1}
  {\emph {\bibinfo {booktitle} {{Nb$_{3}$ Sn Accelerator Magnets}}}},\ \bibinfo
  {editor} {edited by\ \bibinfo {editor} {\bibfnamefont {D.}~\bibnamefont
  {Schoerling}}\ and\ \bibinfo {editor} {\bibfnamefont {A.~V.}\ \bibnamefont
  {Zlobin}}}\ (\bibinfo {year} {2019})\ pp.\ \bibinfo {pages}
  {3--22}\BibitemShut {NoStop}%
\bibitem [{\citenamefont {{L. Bottura \textit{et al}.}}(2018)}]{bottura18}%
  \BibitemOpen
  \bibfield  {author} {\bibinfo {author} {\bibnamefont {{L. Bottura \textit{et
  al}.}}},\ }\bibfield  {title} {\bibinfo {title} {{Expected performance of 11T
  and MB dipoles considering the cooling performance}},\ }\bibfield  {journal}
  {\bibinfo  {journal} {Presentation at 8th HL-LHC collaboration meeting, CERN,
  Geneva Switzerland}\ }\href
  {https://doi.org/https://indico.cern.ch/event/742082/contributions/3085134/}
  {https://indico.cern.ch/event/742082/contributions/3085134/} (\bibinfo {year}
  {2018})\BibitemShut {NoStop}%
\bibitem [{\citenamefont {{C. Bahamonde Castro \textit{et
  al}.}}(2018)}]{anton-et-all-on-TCLD-losses}%
  \BibitemOpen
  \bibfield  {author} {\bibinfo {author} {\bibnamefont {{C. Bahamonde Castro
  \textit{et al}.}}},\ }\bibfield  {title} {\bibinfo {title} {{Energy
  deposition from collimation losses in the DS region at P7}},\ }\bibfield
  {journal} {\bibinfo  {journal} {Presentation at 8th HL-LHC collaboration
  meeting, CERN, Geneva Switzerland}\ }\href
  {https://doi.org/https://indico.cern.ch/event/742082/contributions/3085132/}
  {https://indico.cern.ch/event/742082/contributions/3085132/} (\bibinfo {year}
  {2018})\BibitemShut {NoStop}%
\bibitem [{\citenamefont {Carra}(2013)}]{TCT_fede}%
  \BibitemOpen
  \bibfield  {author} {\bibinfo {author} {\bibfnamefont {F.}~\bibnamefont
  {Carra}},\ }\bibfield  {title} {\bibinfo {title} {{Summary of calculations
  performed on TCTP collimators}},\ }\bibfield  {journal} {\bibinfo  {journal}
  {CERN EDMS 1212639, LHC-TC-ER-0001}\ }\href
  {https://doi.org/{https://edms.cern.ch/document/1212639}}
  {{https://edms.cern.ch/document/1212639}} (\bibinfo {year}
  {2013})\BibitemShut {NoStop}%
\bibitem [{\citenamefont {Mirarchi}(2015)}]{daniele-thesis}%
  \BibitemOpen
  \bibfield  {author} {\bibinfo {author} {\bibfnamefont {D.}~\bibnamefont
  {Mirarchi}},\ }\emph {\bibinfo {title} {{Crystal Collimation for LHC}}},\
  \href {http://cds.cern.ch/record/2036210} {Ph.D. thesis},\ \bibinfo  {school}
  {Imperial College, London} (\bibinfo {year} {2015})\BibitemShut {NoStop}%
\bibitem [{\citenamefont {{W. Scandale \textit{et al.}}}(2016)}]{scandale16}%
  \BibitemOpen
  \bibfield  {author} {\bibinfo {author} {\bibnamefont {{W. Scandale \textit{et
  al.}}}},\ }\bibfield  {title} {\bibinfo {title} {Observation of channeling
  for 6500 gev/c protons in the crystal assisted collimation setup for lhc},\
  }\href {https://doi.org/https://doi.org/10.1016/j.physletb.2016.05.004}
  {\bibfield  {journal} {\bibinfo  {journal} {Physics Letters B}\ }\textbf
  {\bibinfo {volume} {758}},\ \bibinfo {pages} {129 } (\bibinfo {year}
  {2016})}\BibitemShut {NoStop}%
\bibitem [{\citenamefont {Mirarchi}\ \emph {et~al.}(2017)\citenamefont
  {Mirarchi}, \citenamefont {Hall}, \citenamefont {Redaelli},\ and\
  \citenamefont {Scandale}}]{Mirarchi2017_crystals}%
  \BibitemOpen
  \bibfield  {author} {\bibinfo {author} {\bibfnamefont {D.}~\bibnamefont
  {Mirarchi}}, \bibinfo {author} {\bibfnamefont {G.}~\bibnamefont {Hall}},
  \bibinfo {author} {\bibfnamefont {S.}~\bibnamefont {Redaelli}},\ and\
  \bibinfo {author} {\bibfnamefont {W.}~\bibnamefont {Scandale}},\ }\bibfield
  {title} {\bibinfo {title} {Design and implementation of a crystal collimation
  test stand at the large hadron collider},\ }\href
  {https://doi.org/10.1140/epjc/s10052-017-4985-4} {\bibfield  {journal}
  {\bibinfo  {journal} {The European Physical Journal C}\ }\textbf {\bibinfo
  {volume} {77}},\ \bibinfo {pages} {424} (\bibinfo {year} {2017})}\BibitemShut
  {NoStop}%
\bibitem [{\citenamefont {D'Andrea}\ \emph {et~al.}(2019)\citenamefont
  {D'Andrea}, \citenamefont {Mirarchi}, \citenamefont {Redaelli}, \citenamefont
  {Fomin}, \citenamefont {Belli}, \citenamefont {Salvachua~Ferrando},
  \citenamefont {Nevay}, \citenamefont {Rossi}, \citenamefont {Scandale},
  \citenamefont {Montesano}, \citenamefont {Galluccio}, \citenamefont
  {Serrano~Galvez}, \citenamefont {Dionisio~Barreto}, \citenamefont {Butcher},\
  and\ \citenamefont {Lamas~Garcia}}]{D'Andrea:2678781}%
  \BibitemOpen
  \bibfield  {author} {\bibinfo {author} {\bibfnamefont {M.}~\bibnamefont
  {D'Andrea}}, \bibinfo {author} {\bibfnamefont {D.}~\bibnamefont {Mirarchi}},
  \bibinfo {author} {\bibfnamefont {S.}~\bibnamefont {Redaelli}}, \bibinfo
  {author} {\bibfnamefont {A.}~\bibnamefont {Fomin}}, \bibinfo {author}
  {\bibfnamefont {E.}~\bibnamefont {Belli}}, \bibinfo {author} {\bibfnamefont
  {B.}~\bibnamefont {Salvachua~Ferrando}}, \bibinfo {author} {\bibfnamefont
  {L.}~\bibnamefont {Nevay}}, \bibinfo {author} {\bibfnamefont
  {R.}~\bibnamefont {Rossi}}, \bibinfo {author} {\bibfnamefont
  {W.}~\bibnamefont {Scandale}}, \bibinfo {author} {\bibfnamefont
  {S.}~\bibnamefont {Montesano}}, \bibinfo {author} {\bibfnamefont
  {F.}~\bibnamefont {Galluccio}}, \bibinfo {author} {\bibfnamefont
  {P.}~\bibnamefont {Serrano~Galvez}}, \bibinfo {author} {\bibfnamefont
  {C.}~\bibnamefont {Dionisio~Barreto}}, \bibinfo {author} {\bibfnamefont
  {M.}~\bibnamefont {Butcher}},\ and\ \bibinfo {author} {\bibfnamefont
  {I.}~\bibnamefont {Lamas~Garcia}},\ }\bibfield  {title} {\bibinfo {title}
  {{Crystal Collimation Tests with Pb Ion Beams}},\ }\href
  {http://cds.cern.ch/record/2678781} {\bibfield  {journal} {\bibinfo
  {journal} {CERN-ACC-NOTE-2019-0024}\ } (\bibinfo {year} {2019})}\BibitemShut
  {NoStop}%
\bibitem [{\citenamefont {{S. R. Klein}}(2001)}]{klein01}%
  \BibitemOpen
  \bibfield  {author} {\bibinfo {author} {\bibnamefont {{S. R. Klein}}},\
  }\bibfield  {title} {\bibinfo {title} {Localized beampipe heating due to e-
  capture and nuclear excitation in heavy ion colliders},\ }\href@noop {}
  {\bibfield  {journal} {\bibinfo  {journal} {Nucl. Inst. \& Methods A}\
  }\textbf {\bibinfo {volume} {459}},\ \bibinfo {pages} {51} (\bibinfo {year}
  {2001})}\BibitemShut {NoStop}%
\bibitem [{\citenamefont {Bruce}\ \emph {et~al.}(2007)\citenamefont {Bruce},
  \citenamefont {Jowett}, \citenamefont {Gilardoni}, \citenamefont {Drees},
  \citenamefont {Fischer}, \citenamefont {Tepikian},\ and\ \citenamefont
  {Klein}}]{prl07}%
  \BibitemOpen
  \bibfield  {author} {\bibinfo {author} {\bibfnamefont {R.}~\bibnamefont
  {Bruce}}, \bibinfo {author} {\bibfnamefont {J.~M.}\ \bibnamefont {Jowett}},
  \bibinfo {author} {\bibfnamefont {S.}~\bibnamefont {Gilardoni}}, \bibinfo
  {author} {\bibfnamefont {A.}~\bibnamefont {Drees}}, \bibinfo {author}
  {\bibfnamefont {W.}~\bibnamefont {Fischer}}, \bibinfo {author} {\bibfnamefont
  {S.}~\bibnamefont {Tepikian}},\ and\ \bibinfo {author} {\bibfnamefont
  {S.~R.}\ \bibnamefont {Klein}},\ }\bibfield  {title} {\bibinfo {title}
  {Observations of beam losses due to bound-free pair production in a heavy-ion
  collider},\ }\href {https://doi.org/10.1103/PhysRevLett.99.144801} {\bibfield
   {journal} {\bibinfo  {journal} {Phys. Rev. Letters}\ }\textbf {\bibinfo
  {volume} {99}},\ \bibinfo {pages} {144801} (\bibinfo {year}
  {2007})}\BibitemShut {NoStop}%
\bibitem [{\citenamefont {Bruce}\ \emph {et~al.}(2009)\citenamefont {Bruce},
  \citenamefont {Bocian}, \citenamefont {Gilardoni},\ and\ \citenamefont
  {Jowett}}]{prstabBFPP09}%
  \BibitemOpen
  \bibfield  {author} {\bibinfo {author} {\bibfnamefont {R.}~\bibnamefont
  {Bruce}}, \bibinfo {author} {\bibfnamefont {D.}~\bibnamefont {Bocian}},
  \bibinfo {author} {\bibfnamefont {S.}~\bibnamefont {Gilardoni}},\ and\
  \bibinfo {author} {\bibfnamefont {J.~M.}\ \bibnamefont {Jowett}},\ }\bibfield
   {title} {\bibinfo {title} {{Beam losses from ultraperipheral nuclear
  collisions between Pb ions in the Large Hadron Collider and their
  alleviation}},\ }\href {https://doi.org/10.1103/PhysRevSTAB.12.071002}
  {\bibfield  {journal} {\bibinfo  {journal} {Phys. Rev. ST Accel. Beams}\
  }\textbf {\bibinfo {volume} {12}},\ \bibinfo {pages} {071002} (\bibinfo
  {year} {2009})}\BibitemShut {NoStop}%
\bibitem [{\citenamefont {Schaumann}\ \emph {et~al.}(2016)\citenamefont
  {Schaumann} \emph {et~al.}}]{schaumann16_md_BFPPquench}%
  \BibitemOpen
  \bibfield  {author} {\bibinfo {author} {\bibfnamefont {M.}~\bibnamefont
  {Schaumann}} \emph {et~al.},\ }\bibfield  {title} {\bibinfo {title} {{LHC
  BFPP Quench Test with Ions (2015)}},\ }\href@noop {} {\bibfield  {journal}
  {\bibinfo  {journal} {CERN-ACC-NOTE-2016-0024}\ } (\bibinfo {year}
  {2016})}\BibitemShut {NoStop}%
\bibitem [{\citenamefont {{J.M. Jowett \textit{et
  al.}}}(2016)}]{jowett16_ipac_bfpp}%
  \BibitemOpen
  \bibfield  {author} {\bibinfo {author} {\bibnamefont {{J.M. Jowett \textit{et
  al.}}}},\ }\bibfield  {title} {\bibinfo {title} {{Bound-free pair production
  in LHC Pb-Pb operation at 6.37 Z TeV per beam}},\ }\href@noop {} {\bibfield
  {journal} {\bibinfo  {journal} {Proceedings of the International Particle
  Accelerator Conference 2016, Busan, Korea}\ ,\ \bibinfo {pages} {1497}}
  (\bibinfo {year} {2016})}\BibitemShut {NoStop}%
\end{thebibliography}%
\null

\end{document}